# A unified multiscale 3D printer combining single-photon Tomographic Volumetric Additive Manufacturing and Two-Photon Polymerization

Buse Unlu[1,*], Felix Wechsler[1], Ye Pu[1], Christophe Moser[1]

[1]Laboratory of Applied Photonics Devices, School of Engineering, Ecole Polytechnique Fédérale de Lausanne, CH-1015,
Lausanne, Switzerland

Corresponding author: buse.unlu@epfl.ch

## Abstract

Single-photon polymerization ensures rapid photopolymerization of centimeter-scale structures with features on the order of tens of micrometers, whereas two-photon polymerization provides sub-micrometer features at sub-millimeter scales. Existing hybrid approaches combining these techniques typically rely on stitched or layer-by-layer fabrication and often require separate printing platforms, making mesoscale manufacturing time-consuming. Here, we introduce a hybrid unified 3D printer that leverages the complementary strengths of both printing mechanisms to bridge this scale resolution-fabrication time gap.
We propose integrating two-photon polymerization (2PP) for high-resolution, localized spatial control with single-photon Tomographic Volumetric Additive Manufacturing (TVAM) for enabling rapid, high-throughput 3D fabrication. In this approach, TVAM first forms millimeter-scale volumetric structures attached on a glass rod, via overprinting, which is then accessible, on the same platform, for subsequent high-resolution 2PP. Without needing to change the photoresin or introducing intermediate post-processing steps, we proceed to demonstrate finely printed structures via 2PP, fabricated both inside (embedded within) and on the surface of the millimeter-scale 3D objects printed with TVAM. Here, TVAM contributes in two distinct ways: by generating a pre-polymerized volume that facilitates subsequent 2PP, and by directly driving layer-less volumetric polymerization in designated regions within seconds. We experimentally demonstrate that this dual-mode strategy provides a mesoscale approach spanning three orders of magnitude in scale for rapid fabrication of millimeter-scale structures featuring 830 nm details. For applications such as micro-optics, biomedical scaffolds and tissue engineering, tens-of-micrometer features are sufficient across the majority of the volume, with higher resolution confined to localized functional regions.

## 1. Introduction

Light-based additive manufacturing has emerged as a powerful fabrication technique, producing complex three-dimensional (3D) structures with high speed, high spatial precision and broad material versatility and compatibility, which are challenging to achieve using conventional manufacturing approaches[1–3]. However, a fundamental challenge in optical 3D printing is the trade-off between spatial resolution and fabrication throughput. Techniques capable of achieving nanometer-scale precision typically rely on serial writing processes that limit fabrication speed and build volume, whereas high-throughput optical methods generally sacrifice spatial resolution[2,4–6]. As a result, the fabrication of mesoscale structures spanning several orders of magnitude in length scale from sub-micrometer features to millimeter or centimeter dimensions remains difficult and open.

Several hybrid and adaptive manufacturing strategies have been proposed to bridge this resolution-throughput gap[7–17]. Two-photon polymerization techniques can achieve extremely high spatial resolution, with typical feature sizes in the sub-micrometer range (hundreds of nanometers) under high-NA objective focusing and reaching feature sizes down to tens of nanometers in specialized configurations such as near-threshold exposure or STED-inspired 2PP with optimized resin formulations[18]. However, because microscope objectives provide a limited field of view (tens to hundreds of micrometers), centimeter-scale structures must be fabricated by stitching together multiple scanning fields using mechanical stages. Consequently, fabrication times typically extend to several hours[2,4]. Even stitch-less mesoscale implementations require synchronized galvo and stage scanners and still take hours to fabricate centimeter-scale structures[7]. Micro-stereolithography (µSLA) represents another approach for achieving micrometer-scale resolution using single-photon SLA with microscope objectives. In this case as well, mechanical scanning or lens arrays are required to enlarge the printable surface area, resulting in fabrication times ranging from tens of minutes to several hours[8,9]. Digital Light Processing (DLP) systems enable faster fabrication over larger build areas, but the achievable feature size remains limited to tens of micrometers[19].

Hybrid systems combining conventional single-photon lithography techniques, namely DLP and µSLA, have been explored for producing mesoscale structures; however, they rely on layer-by-layer printing, which inherently causes longer printing time typically ranging from tens of minutes to hours[10]. In most reported implementations, mesoscale fabrication relies on sequential multi-platform workflows, where a structure is first fabricated using one system (for example, DLP or SLA) and then transferred to another system, such as two-photon polymerization system, for high-resolution processing[14–17]. Integrated systems combining single-photon lithography techniques (such as DLP or SLA) with two-photon polymerization remain relatively rare[11–13]. Ref.[12,13] have not reported experimental results to our knowledge. Supplementary Table 1 in Supplementary Materials summarizes the state-of-the-art in optical based mesoscale manufacturing techniques.

Two optical fabrication techniques illustrate complementary capabilities that could potentially address this challenge: tomographic volumetric additive manufacturing (TVAM) and two-photon polymerization (2PP). TVAM enables rapid, volumetric fabrication of structures[20–22], whereas 2PP allows precise fabrication of intricate microscale features[23,24].

TVAM is a single-photon[20,22,25] absorption 3D-printing technique that relies on the projection of tomographic light patterns onto a rotating photocurable resin container. A photoinitiator inside the resin absorbs the projected light and triggers a photopolymerization reaction. Due to the rotation of the resin container during illumination, the absorbed light dose can be spatially controlled throughout the entire volume using tomographic projections, and once a local dose threshold is reached, a stable polymerization network is formed[2,26], building the final object. This layer-less fabrication approach enables extremely rapid volumetric printing, with typical fabrication times between a few seconds and several tens of seconds for prints reach up to several centimeters in extent[27]. However, the achievable spatial resolution remains limited, with minimum feature sizes typically on the order of tens of micrometers[28].

Two-photon polymerization, in contrast to TVAM, is a nonlinear process in which a molecule absorbs two photons simultaneously. This effect is more pronounced within the focal point of a tightly focused femtosecond (fs) pulsed laser[29–31]. Because the probability of two-photon absorption (2PA) depends nonlinearly on the local light intensity, polymerization occurs only within the focal volume of the beam. By scanning the fs laser focal spot throughout the resin, individual voxels can be sequentially written to construct 3D structure[30,32]. This highly localized confinement in the volume enables extremely high-precision, with a feature sizes below 100 nm reported in the literature[33,34]. However, because the fabrication relies on serial voxel-by-voxel scanning, and requires high peak intensities due to the low 2PA cross-section[29,35], the overall fabrication throughput remains limited.

Here, we demonstrate a hybrid additive manufacturing approach that combines layer-less TVAM with high-resolution 2PP within a single experimental platform. In our system, TVAM is first used to rapidly fabricate millimeter-scale structures within seconds. Because TVAM does not rely on layer-by-layer fabrication and does not require support structures, it enables flexible overprinting on existing elements[20,36,37] and provides access to target surfaces for subsequent high-resolution processing via 2PP. A glass rod serves as an overprinting platform, and localized polymerization is initiated on top of the glass rod. Without removing the sample or performing intermediate post-processing, the structure is further processed by 2PP to directly fabricate micrometer- and sub-micrometer-scale features both at the surface and within the volumetrically polymerized, millimeter-scale structures.

We also introduce the concept of a “volume feature ratio” as a general measure of print complexity, defined as the ratio between the volume printed by TVAM and the volume printed via 2PP. Building on this concept, we further propose using the time rate of this ratio as a reference standard for throughput benchmarking of hybrid systems.

Using this approach, we experimentally demonstrate hybrid structures spanning three orders of magnitude in length scale. In our experiments, a 1 $mm^3$ volume was fabricated using TVAM in approximately 12 seconds, while subsequent two-photon features with an 830 nm feature size were written in approximately three minutes. This hybrid strategy therefore enables rapid fabrication of mesoscale structures combining centimeter-scale throughput with sub-micrometer resolution. Such a capability is particularly attractive for applications requiring rapid fabrication of millimeter-scale devices with localized nanoscale functionality, such as bioengineered scaffolds and microfluidic devices. Moreover, the distinct refractive indices of the regions printed by 2PP and TVAM[38–40] can be exploited for light guiding, enabling micro-optical components such as waveguides, microlenses and other integrated photonic functionalities.

Overall, this work provides the first demonstration of a unified, single-platform 3D printing system in which TVAM offers layer-less, rapid volumetric fabrication, while 2PP adds localized surface and embedded features within the same photoresin and without intermediate post-processing between the two printing steps, thereby bridging the resolution-throughput trade-off in light-based additive manufacturing. Table 1 highlights key differences between representative hybrid approaches including 2PP discussed above and the TVAM–2PP system introduced in this work.

**Table 1.** Representative hybrid optical 3D-printing approaches and key features compared to this work.

| Approach | Single platform | Volumetric (tomographic) printing | 2PP on surface of bulk | 2PP embedded in bulk | Sample transfer between methods |
|---|---|---|---|---|---|
| LCM/DLP+2PP[15] | No | No | Yes | No | Yes |
| LCD/DLP+2PP[16] | No | No | Yes | No | Yes |
| DLP+2PP (patent[12], press release[13]) | Yes | No | Potentially yes (no result) | No (no result) | No |
| **This work: TVAM+2PP** | **Yes** | **Yes (TVAM)** | **Yes** | **Yes** | **No** |

## 2. Materials and Methods

### 2.1 Experimental setup

We first constructed TVAM and 2PP setups separately (comprehensive sketches of the experimental setups are depicted in Supplementary Material Figure S1 and Figure S2, respectively) and then integrated them to form the hybrid platform shown in Figure 1. A global Cartesian coordinate system (x, y, z) is defined for the integrated setup, where the 2PP fs laser beam propagates along the z-axis and the TVAM illumination is projected in the yz-plane along the x-axis direction. A real photograph of the printing region in the combined TVAM and 2PP setup is provided in Supplementary Material Figure S3. For the TVAM setup, a continuous-wave (CW) laser diode at 405 nm is coupled into a square-core multimode fiber that illuminates the aperture of DMD (V-7000 VIS, Vialux). The generated DMD patterns, described in Section 2.2, are demagnified by a factor of two using a pair of achromatic doublets (L1 and L2 with 300 mm and 150 mm focal lengths, respectively) and imaged in yz-plane into a quartz glass vial (refractive index 1.4696, inner diameter 4.30 mm) containing the photocurable resin. A glass rod (refractive index 1.4814, diameter 1 mm) is dipped into the resin for overprinting and served as a printing platform. Ray-tracing simulations were performed to model the aberration of the beam spot in the printing region using Zemax OpticStudio, yielding a root-mean-square (RMS) point-spread-function (PSF) diameter of 6.8 µm (see Supplementary Material Figure S4). The TVAM printing region is monitored by a camera (C1), using a pair of achromatic doublet lenses (L3, 75 mm focal length). The minimum printable size is 38.3 ± 9.37 µm ( details are provided in the section “Feature size analysis of TVAM system” in Supplementary Materials).

For 2PP, a mode-locked Ti:Sapphire femtosecond pulsed laser beam at 780 nm (repetition rate 80 MHz, pulse duration 70 fs) is directed onto a phase-only liquid crystal on silicon (LCOS) Spatial Light Modulator (SLM, PLUTO-2.1, Holoeye), which provides lateral scanning (in xy-plane) of the fs laser spot over the printing region by changing the period of a blazed grating. The beam is then focused on the TVAM-printed object using a 20×/0.45 numerical aperture (NA) microscope objective (Zeiss A-Plan 20×, NA = 0.45, WD = 0.46 mm). The 3D structure is constructed by 2PP in a voxel-by-voxel manner by digitally scanning the fs laser beam in the xy-plane using the SLM, and in a layer-by-layer manner by translating the glass rod along the z-axis with a motorized translation stage (PT1/M-Z9, Thorlabs). The motorized stage has a minimum incremental movement of 0.2 µm (controlled by MATLAB software), which determines the precision of the rod position. The 2PP printing process is monitored via the camera (C2) through the combination of the objective lens, a long-pass dichroic mirror (cut-on wavelength at 699 nm) and a tube lens (TL, 40 mm focal length). The same imaging system is used to ensure that the fs laser beam is precisely focused onto the printing surface by monitoring its specular reflection. A detailed description of the complete 2PP system and procedure is provided in Supplementary Materials S2 and mainly in our previous publication[41].

Both the glass vial and the glass rod are mounted on separate motorized rotary stages (X-RSW60C, Zaber) respectively, each placed on a one-axis motorized translation stage that provides z-axis motion. The two rotary stages are synchronized to rotate simultaneously via an external controller (X-MCC2, Zaber), enabling overprinting of TVAM structures onto the glass rod. To ensure time synchronization between all devices, the Zaber rotary stages serve as a clock. It provides regular synchronization signals to an Arduino Nano Every, which then forwards these signals to the VialuxDMD display. Based on rotation speed and the number of patterns, the DMD displays each pattern for a certain amount of time.

The fabrication workflow begins by calibrating and recording the DMD pattern projection (yz-plane) with respect to the bottom of the rod using the TVAM camera, C1. After the vial is filled, the fs beam focus position with respect to the glass rod is recorded by monitoring the specular reflection of the fs laser focus from the rod surface using the camera C2 before overprinting by TVAM. Hence there is a direct calibrated correspondence between the fs focus and the TVAM-printed pattern in the z-axis direction. The rod is then repositioned with respect to the DMD light patterns, and TVAM overprinting is performed to fabricate the millimeter-scale structure on the rod. Following TVAM, the rod is translated downward toward the objective working distance, and the TVAM-overprinted structure is repositioned into the 2PP focal plane by monitoring the specular reflection of the fs laser focus from the TVAM surface with camera C2. We use a low power (6 mW) for the fs beam during this calibration to avoid polymerization. Once the correct position is identified, the fs laser power is set to the writing value and high-resolution 2PP features are fabricated either on the surface of or within the TVAM-printed volume. Then, the residual resin surrounding the structures was rinsed by isopropanol alcohol (IPA) and post-cured under a UV lamp to obtain the final hybrid structure.

In summary, the workflow consists of rod alignment with respect to the TVAM light patterns, reference focal-position recording, TVAM overprinting, repositioning into the 2PP focal plane, localized 2PP writing, and final development/post-curing.

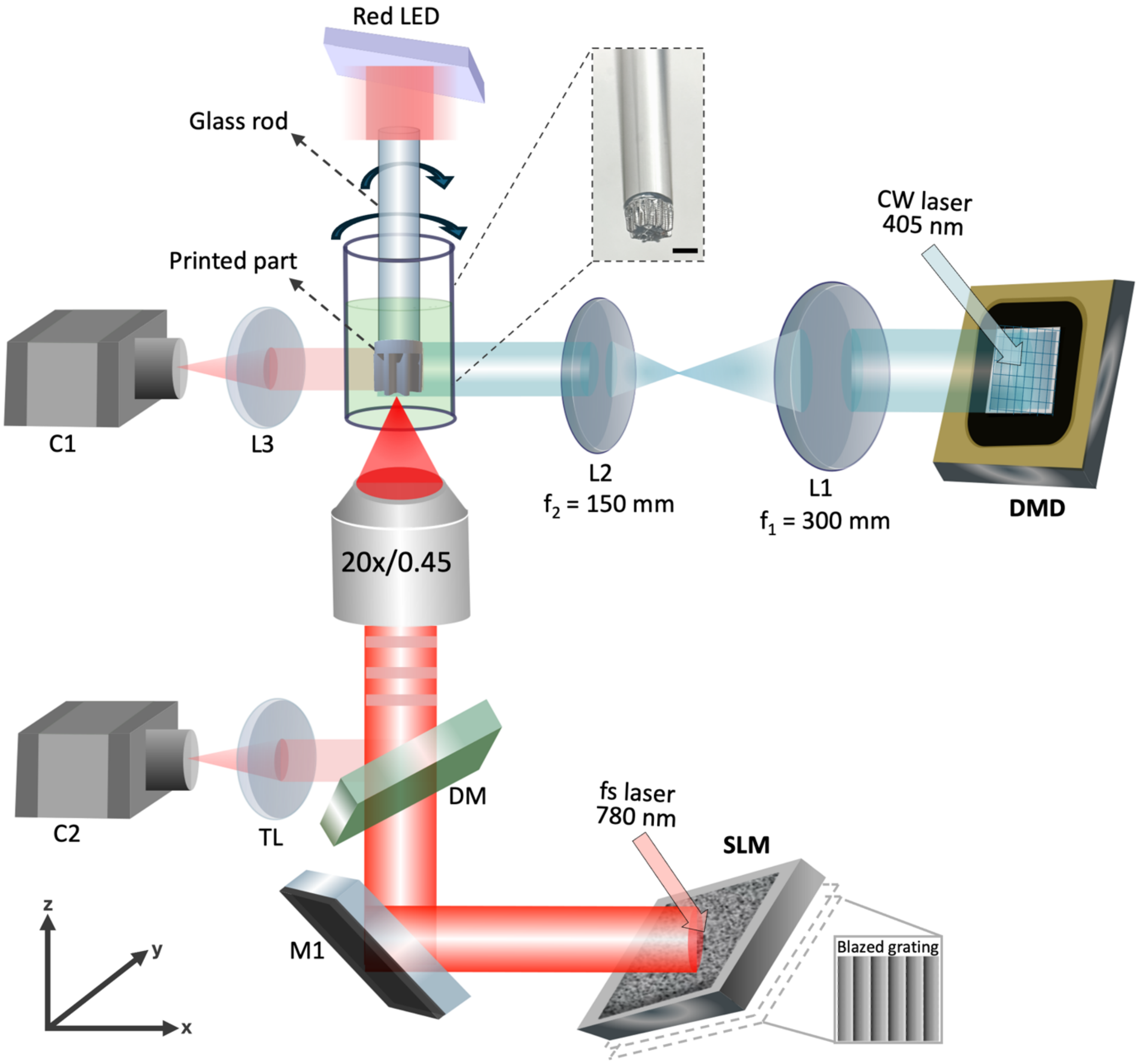

**Figure 1.** Schematic illustration of the experimental setup for the multiscale 3D printer combining single-photon Tomographic Volumetric Manufacturing (illumination in yz-plane) and Two-Photon Polymerization (fs laser propagation along the z-axis with voxel-by-voxel fabrication in the xy-plane): lens, C: camera, M: mirror, DM: dichroic mirror, TL: tube lens. The inset shows a real photograph of a TVAM-overprinted gear structure on a glass rod. The scale bar is 500 μm.

## 2.2 Pattern optimization

In this part, we provide details about the pattern optimization process for the tomographic projections in TVAM. The patterns are calculated with the open-source software Dr.TVAM[36,42] which models ray optical effects such as refraction, absorption, and reflection. Dr.TVAM steps each DMD pixel as a geometrical ray through a discretized voxel grid. At each step, some of the carried light intensity is absorbed by the voxel according to Beer-Lambert's law, and the ray is attenuated. This procedure is repeated until the ray leaves the vial. Overall, the absorbed dose is calculated by summing the absorbed doses of each individual ray from the DMD pixels and all angles. In our case, in addition to attenuation, we also have reflection and refraction on the air-glass-resin interfaces. Further, since we overprint a portion of the structures over the glass rod, we additionally need to model how light propagates from the resin into the glass rod. Inside the glass rod, there is no attenuation. However, there is refraction and reflection which is handled by the Dr. TVAM framework physically correctly. We are using the following loss function in our gradient-descent-based optimization

$$\mathcal{L} = \underbrace{w_{in} \cdot \sum_{v \in object} ReLU(t_u - I_v)^2}_{force\ polymerization\ in\ object} + \underbrace{w_{out} \cdot \sum_{v \notin object} ReLU(I_v - t_l)^2}_{prevent\ polymerization\ elsewhere}$$

$$+ \underbrace{w_o \cdot \sum_{v \in object} ReLU(I_v - 1)^2}_{avoid\ over-polymerization} + \underbrace{w_{sparsity} \cdot \sum_{j \in object} \left|P_j\right|^D}_{enforce\ non-sparse\ patterns} \tag{1}$$

In this equation, $I_v$ represents the absorbed intensity in voxel $v$ after the projection of the patterns. $P_j$ denotes the value of the $j$-th pixel of the patterns. The variables $t_u$ and $t_l$ correspond to the upper and lower thresholds, respectively. The weights $w_{in}$, $w_{out}$, $w_o$, and $w_{sparsity}$ are relative coefficients for the respective terms. The parameter $D$ imposes varying penalties for sparse values. The general goal of this loss function is to find a set of patterns such that projected intensity results in an absorbed dose where object voxels cross the polymerization threshold, and void voxels stay below it. In practice, a set of right printing time and printing power needs to be experimentally determined to obtain a correctly cured print. For our experiments, for a gear structure, we set $t_l$= 0.7, $t_u$ = 0.9, $D$ = 4, $w_{in}$= $w_{out}$= $w_0$ = 1, while for two stacked rectangular prisms with lateral through-holes, we chose $w_{sparsity}$= 1.0. The sparsity term allows us to increase the pattern efficiency of our patterns, which reduces printing time and laser power. For the gear, we did not use sparsity tuning and hence $w_{sparsity}$ = 0. The glass rod, on which the elements were printed, was modeled with a refractive index of $n$ = 1.4814 and a diameter of 1 mm. The optimization time on an NVIDIA L40s was around 3 minutes. The full configuration files and meshes can be found at: https://github.com/EPFL-LAPD/A-unified-multiscale-3D-printer-combining-single-photon-TVAM-and-Two-Photon-Polymerization.

Figure 2a demonstrates the projection of optimized tomographic patterns of the prisms (inverted) onto a glass rod dipped into the resin for overprinting, where $\theta_i$ to $\theta_m$ represent examples of optimized light patterns from different angles projected by the DMD in yz-plane. An example of the pattern at 0° is shown in Figure 2b. The final absorbed light dose from the legs of the small prism, after projecting all tomographic patterns, is depicted in Figure 2c.

It is important to note that, although an index-matching bath is commonly used in TVAM, in our case we did not need to employ one but instead relied on refraction correction by Dr. TVAM. This simplifies the experimental procedure, particularly for 2PP, since the objective lens (OBJ) requires a coverglass thickness of 170 µm to ensure proper aberration correction and optimal focusing.

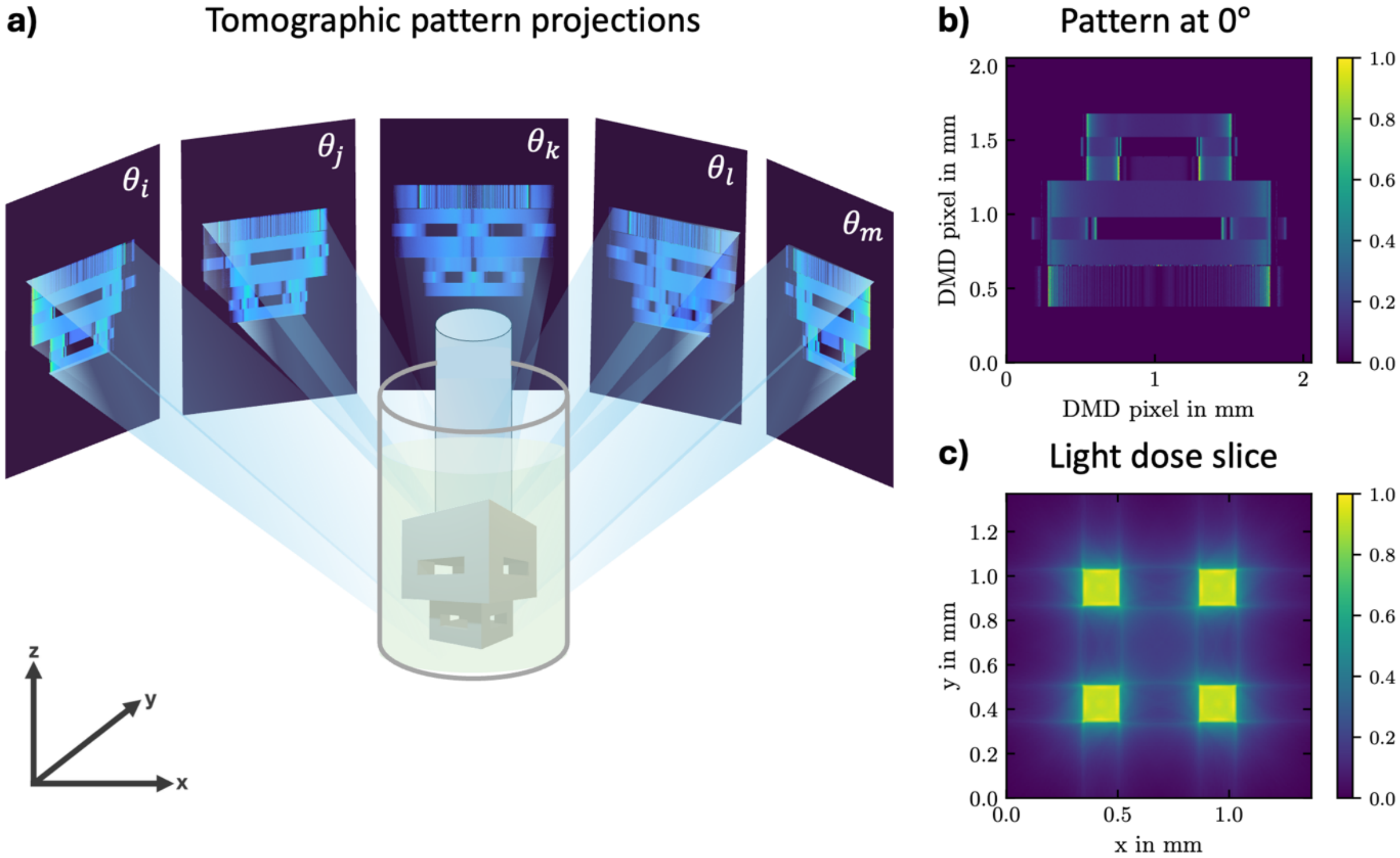

**Figure 2.** Pattern optimization for TVAM. a) Projection of optimized tomographic patterns (increased contrast for visibility), overprinting two stacked rectangular prisms (inverted) onto a glass rod. $\theta_i$ to $\theta_m$ represent examples of optimized light patterns from different angles projected by the DMD in yz-plane. b) An example pattern for the prisms at 0° and c) The final absorbed light dose after projecting all tomographic patterns, a slice taken from the 3D dose of the small prism's legs.

### 2.3 Imaging and characterization

Imaging and characterization were performed using a Zeiss Merlin field-emission SEM and phase-contrast (PC) microscopy. Prior to SEM imaging, the TVAM- and 2PP-fabricated structures were sputter-coated with a 9 nm gold layer using an Alliance-Concept DP 650 magnetron sputtering system. Then, SEM images of the structures were acquired with an accelerating voltage of 3.0 kV, probe current of 100 pA, working distance of 4–6 mm, and using the High-Efficiency Secondary Electron (HE-SE2) detector. The PC microscopy was carried out using a commercial inverted microscope (Olympus IX71) equipped with phase-contrast components, including a condenser annulus and phase-ring objectives with magnifications ranging from 4× to 40×, and a monochrome camera (EO-3212M), enabling visualization of refractive-index variations between TVAM- and 2PP-processed regions.

### 2.4 Materials

A photocurable resin composed of diurethane dimethacrylate (DUDMA, Sigma Aldrich) and poly(ethylene glycol) diacrylate (PEGDA, average Mn 700 g/mol, Sigma Aldrich) in a 4:1 weight ratio is used for fabrication, with ethyl-2,4,6trimethylbenzoylphenylphosphinate (TPO-L, 97 %, Chemie Brunschwig) photoinitiator at a concentration of 43 mM, enabling both single- and two-photon excitation[41,43–46]. The resin composition is mixed using a planetary mixer (Kurabo Mazerustar). The refractive index of the resin was measured to be 1.4827 using a digital refractometer (operating at 589 nm, Reichert AR200). The absorbance of the final resin was calculated to be 4.695, corresponding to an extinction coefficient of 1.081 $mm^{-1}$, as detailed in Supplementary Materials Figure S5. This value represents the highest absorbance at which TVAM still operates reliably for the designed structural geometries and is therefore used in the pattern optimizations.

For post-processing, the overprinted structures were removed from the resin vial and immersed in isopropanol (99 %, Thommen-Furrer AG, Switzerland) for 10 minutes to rinse away the residual uncured resin surrounding the structures. The samples were then detached from the glass rod, placed on a microscope slide, and fully air-dried at room temperature. Finally, the samples were post-cured under a UV lamp (Solis-405C, 100 mW, Thorlabs, 100 $mW/cm^2$ at sample location) for 1 minute to obtain the final hybrid structures.

## 3. Results and discussion

We designed the gear structure shown in Figure 3a using a computer-aided design (CAD) software for printing by TVAM. The height and width of the gear were set to 1.1 mm, and the portion to be overprinted on the glass rod was 225 μm in height (z-axis). For 2PP, two-dimensional (2D) image of the QR code of our group website and its letters, as shown in Figure 3b, served as patterns for scanning by the SLM in xy-plane. In the experiment, before placing the resin vial, the DMD light patterns of the gear structure were sent, and the position of the rod was aligned to the designed location using a motorized linear stage in z-axis controlled by MATLAB software, where the alignment was verified using the camera C1 in Figure 1. Also, the location of rod where the fs laser beam is focused (working distance) was recorded for use in 2PP. The vial and glass rod were then mounted on rotary stages and positioned at the center of the DMD image plane. To ensure a uniform rotational state of the viscous resin, the glass vial was rotated for three full turns prior to DMD illumination. This pre-rotation prevents a rotational velocity mismatch within the resin that would otherwise lead to curved gear teeth, as illustrated in Figure S9.

The optimized 2D patterns described in Section 2.2 were projected onto the photocurable resin with a light dose of 45.3 $mJ/cm^3$ (the calculation is provided in Supplementary Materials). Within 12 seconds, the gear was overprinted onto the glass rod. Without removing the gear-overprinted rod or applying any post-processing steps, the rod was simply moved downward using the motorized stage to reach the working distance of the objective lens (in Figure 1), thereby initiating 2PP. The hatching distance (distance between the voxels, lateral overlap in xy-plane) was experimentally set to 760 nm along both the x- and y-axes for the present optical configuration to provide sufficient overlap between neighboring written voxels while maintaining a practical writing time, while the slicing distance (distance between two consecutive layers) was 0.8 μm. We printed a two-layer QR code on the surface of the gear using an average IR fs laser power of 13 mW and an exposure time of 50 ms per laser spot. After the printing process, the final print was removed from the glass rod. Then, the residual resin surrounding the structures was rinsed in IPA for 10 minutes and UV post-cured for 1 minute. Figure 3c shows the scanning electron microscope (SEM) image of the final object, a millimeter-scale TVAM-overprinted gear featuring sub-micrometer details fabricated by 2PP. A minimum lateral feature size was 830 nm (detailed analysis is provided in Supplementary Materials S6). Due to the sub-micrometer feature size, insufficient optical contrast and minor defects, the QR code cannot be decoded using a smartphone camera[47]. The minor imperfections likely arise from contributing factors such as residual local monomer conversion inhomogeneity in the resin after TVAM exposure which yields differences in crosslinking density between TVAM and 2PP processed volumes thus generating local shrinkage, stress or swelling effects during the final IPA rinsing and UV post-curing steps[48–50]. These observations indicate that further optimization of the common resin formulation, 2PP parameters and the development protocol could improve the quality of the 2PP-written features in the hybrid platform. In addition, Grayscale exposure strategies for SLM-based 2PP[51], which are compatible with our configuration, also offer a concrete avenue for improving print quality in future implementations.
In a second experiment, by placing the rod off-center relative to the focus of the objective lens, the system enables more structures to be printed by 2PP at multiple locations along the rotational path of the TVAM-overprinted structure. Figure 3d illustrates the sequential 2PP-printed structures, including the QR code and letters, by a sequential rotate-stop-print process of the glass rod using the same printing parameters as in the previous experiment. The image was captured by the system camera C2 (in Figure 1) right after the printing process, before the post-processing steps.
In addition, a further experiment was conducted in which 2PP printing was performed during continuous rod rotation. In this case, instead of scanning the fs laser beam using the SLM, the focused beam with 13 mW average power was kept at a fixed single position. The rod supporting the TVAM-overprinted gear was rotated at an angular speed of 2 °/s, causing the gear surface to move through the fixed focal spot. This rotation effectively served as the writing motion, replacing conventional scanning in the 2PP process and producing a circular trace on the gear surface. Figure 3e presents the resulting 2PP-printed structure, with an approximate diameter of 168 µm, obtained after two consecutive full rotations of the rod, enabling direct printing over an extended area on top of the gear. Because the gear was not perfectly centered with respect to the rotation axis, the printed circular structure appears off-center relative to the gear geometry. In addition, wobble of the rotation system created a distorted circular trace.
While the rotation of the rod is a requirement for TVAM, these experiments show that it can be advantageously adapted for 2PP, thereby increasing the accessible field-of-view (FOV) on the TVAM-printed structure, and extending the versatility and functionality of the multiscale 3D printer setup.

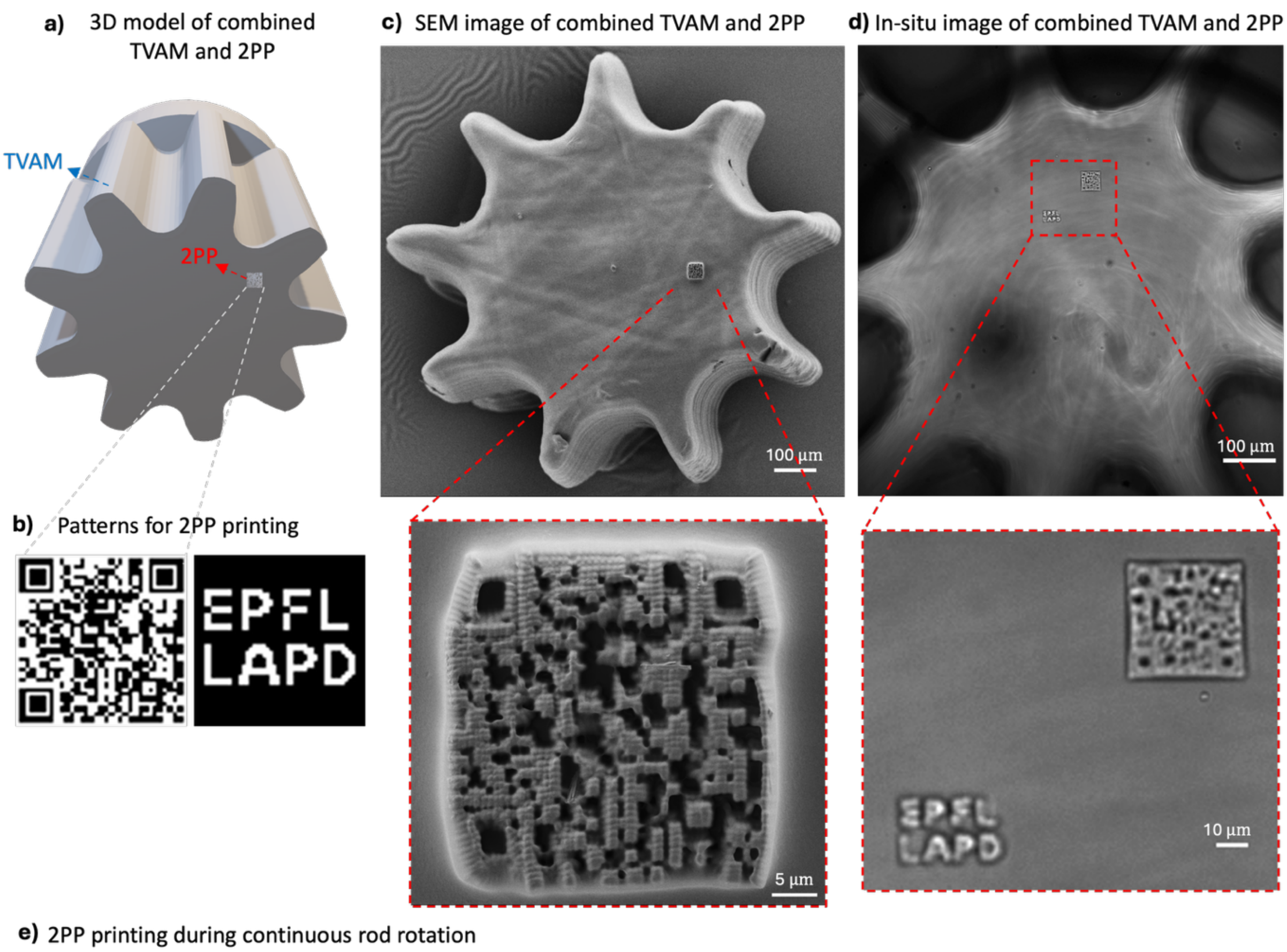

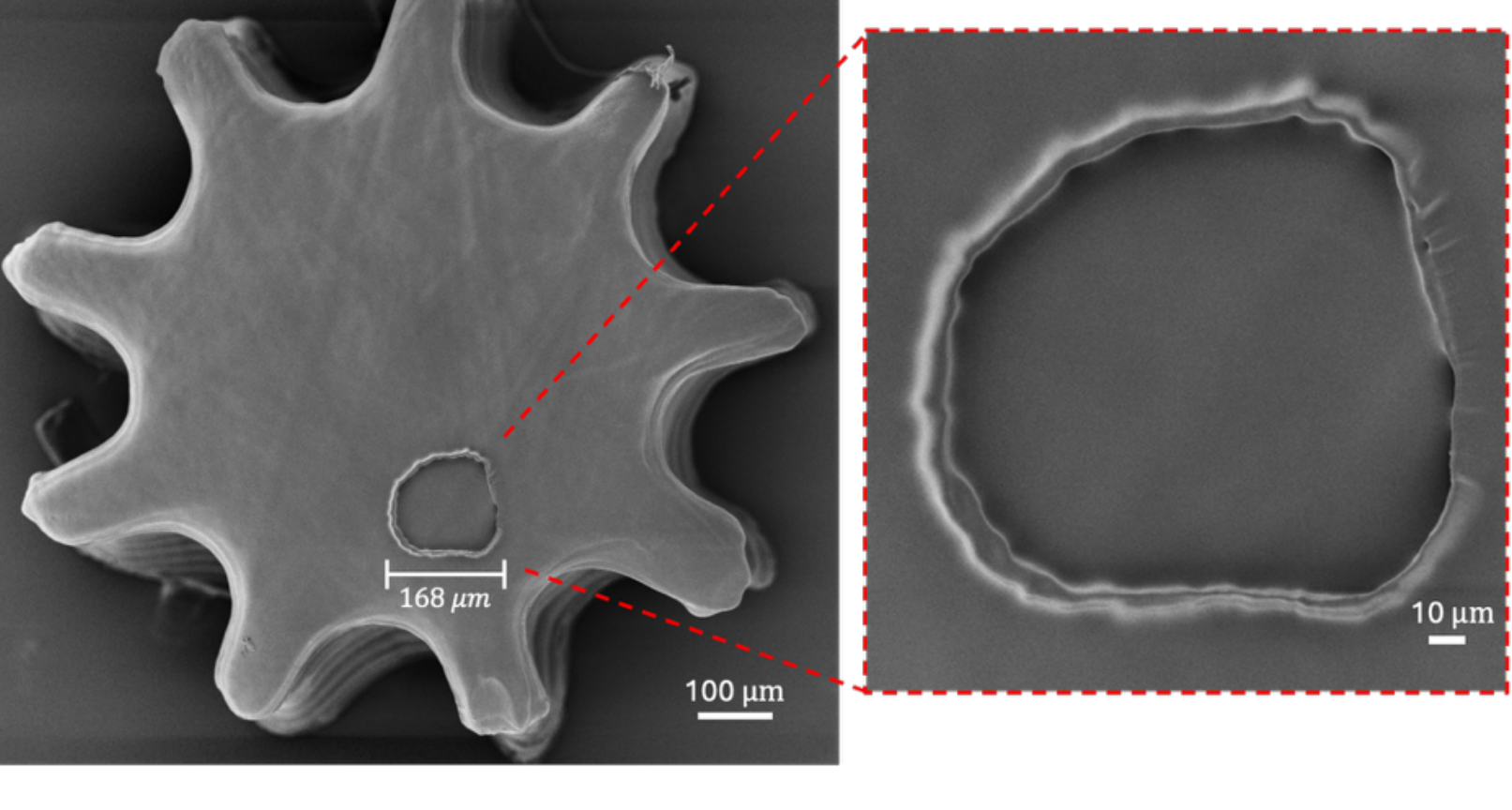

**Figure 3.** Millimeter-scale TVAM-overprinted gear with sub-micrometer details fabricated by 2PP. All images are shown in the xy-plane. a) 3D CAD model for the combination of TVAM (gear) and 2PP (QR code) printing. b) The 2D image patterns for 2PP printing. c) SEM image of the final structure printed by a combination of TVAM and 2PP. The inset shows a magnified view of the QR code printed by 2PP. Scale bars are 100 µm and 5 µm, respectively. d) Camera-captured image of a gear featured by sequential 2PP-printed features (QR code and letters), formed by a sequential rotate-stop-print process of the glass rod. The inset is a magnified view of the 2PP-printed structures. The scale bars are 100 µm and 10 µm, respectively. e) 2PP-printed structure (~168 µm) during the continuous rod rotation on the TVAM-printed gear. The scale bars are 100 µm and 10 µm, respectively.

Following the gear fabrication, a more complex 3D structure shown in Figure 4a was designed. It consists of two stacked rectangular prisms (1.1 x 1.1 x 0.85 mm and 0.68 x 0.68 x 0.45 mm) with lateral through-holes (1.1 x 0.60 x 0.15 and 0.68 x 0.36 x 0.13 mm) for printing with TVAM, and a bridge-like structure with four legs (36 x 36 x 28 µm) and a top hole (15 x 15 x 4 µm) for 2PP. The portion of the bigger rectangular prism to be overprinted on the glass rod (inverted) was 275 µm in height.

The optimized 2D light patterns mentioned in Section 2.2 were projected onto the resin with a light dose of 51.3 mJ/cm$^3$, and within 12 seconds, the structure composed of rectangular prisms was overprinted onto the glass rod. The rod was then moved downward using the motorized stage until it reached the working distance of the objective lens (in Figure 1). At this location, 2PP was initiated. The hatching distance was set to 760 nm along both the x- and y-axes, while the slicing distance was 1.2 µm. The bridge-like structure (23 layers) was printed on the surface of the smaller rectangular prism using an average IR fs laser power of 18 mW and an exposure time of 50 ms per laser spot. After printing, the final structure was removed from the glass rod, and the unsolidified resin was washed out with IPA. Figure 4b shows a digital microscope (VHX-5000, Keyence) image of the final object, which comprises TVAM and 2PP structures printed sequentially on top of each other. A magnified view of the 2PP-printed region is shown in the inset, with a detailed SEM image provided in Figure 4c. As can be observed, in TVAM-overprinted rectangular prisms, corners appear rounded mainly due to optical diffraction and dose accumulation effects, which blur sharp features[52]. In contrast, 2PP achieves sharp edges thanks to non-linear 2PA and highly localized voxel-by-voxel printing.

In the following experiment, we focused the fs laser beam beneath the outer surface of the TVAM-overprinted part to fabricate embedded microstructures. The working distance of the 0.45 NA microscope objective is 460 µm. The microscope glass thickness at the bottom of the vial is 170 µm. Hence the maximum focus distance from the glass is 290 µm. Optical simulation (Zemax OpticStudio) shows that the system comprising the microscope objective, the tube lens and the coverglass thickness does not introduce significant aberrations of the focus spot up to a focus distance of 50 µm along the optical z-axis inside the resin (details in Supplementary Figure S9 of Ref.[41]). Therefore, when the TVAM printed structure is lowered in the vial to touch the glass slide, the maximum focus spot distance inside the TVAM structure is 50 µm. Thus, in an embedded-printing experiment, the 2PP focal spot can in principle be placed up to 50 µm beneath the outer surface of the TVAM-printed prism and scanned within the pre-polymerized bulk. We chose a conservative setting and positioned the fs beam spot approximately 40 µm beneath the outer surface. In this configuration, 2PP locally increases the degree of crosslinking inside the existing TVAM structure. The resulting crosslinking-density differences between the single- and two-photon illumination produces a refractive-index contrast. As expected, the 2PP-printed bridge-like structure is not visible in the SEM images in Figure 4d since it is embedded within the TVAM printed part. However, the embedded structure is clearly observable under a phase contrast (PC) microscope, as shown in Figure 4e. This is because the PC microscope reveals variations in refractive index between the 2PP-modified and TVAM-only regions, which are not detectable in SEM, showing only surface morphology. In contrast to many previous hybrid approaches, where high-resolution features are restricted to exposed surfaces of a pre-printed part, here the pre-polymerized TVAM volume itself acts as a transparent host that can be selectively re-crosslinked by 2PP. This is relevant for applications requiring in-volume photonic functionality. For example, embedded 2PP structures could act as internal photonic elements such as embedded Bragg periodic index modulations for spectral filtering or light coupling inside a transparent host volume[53–55]. In future implementations, this approach could be extended towards integrated photonic elements inspired by holographic waveguide architectures used in near-eye holographic AR displays, where embedded nanoscale gratings and holographic optical elements in thin glass or polymer substrates guide light to the eye, with TVAM defining the macroscopic carrier volume and 2PP writing the embedded high-index or periodic structures (e.g., Bragg-like gratings, couplers, or localized micro-lenses) inside that volume[56,57].

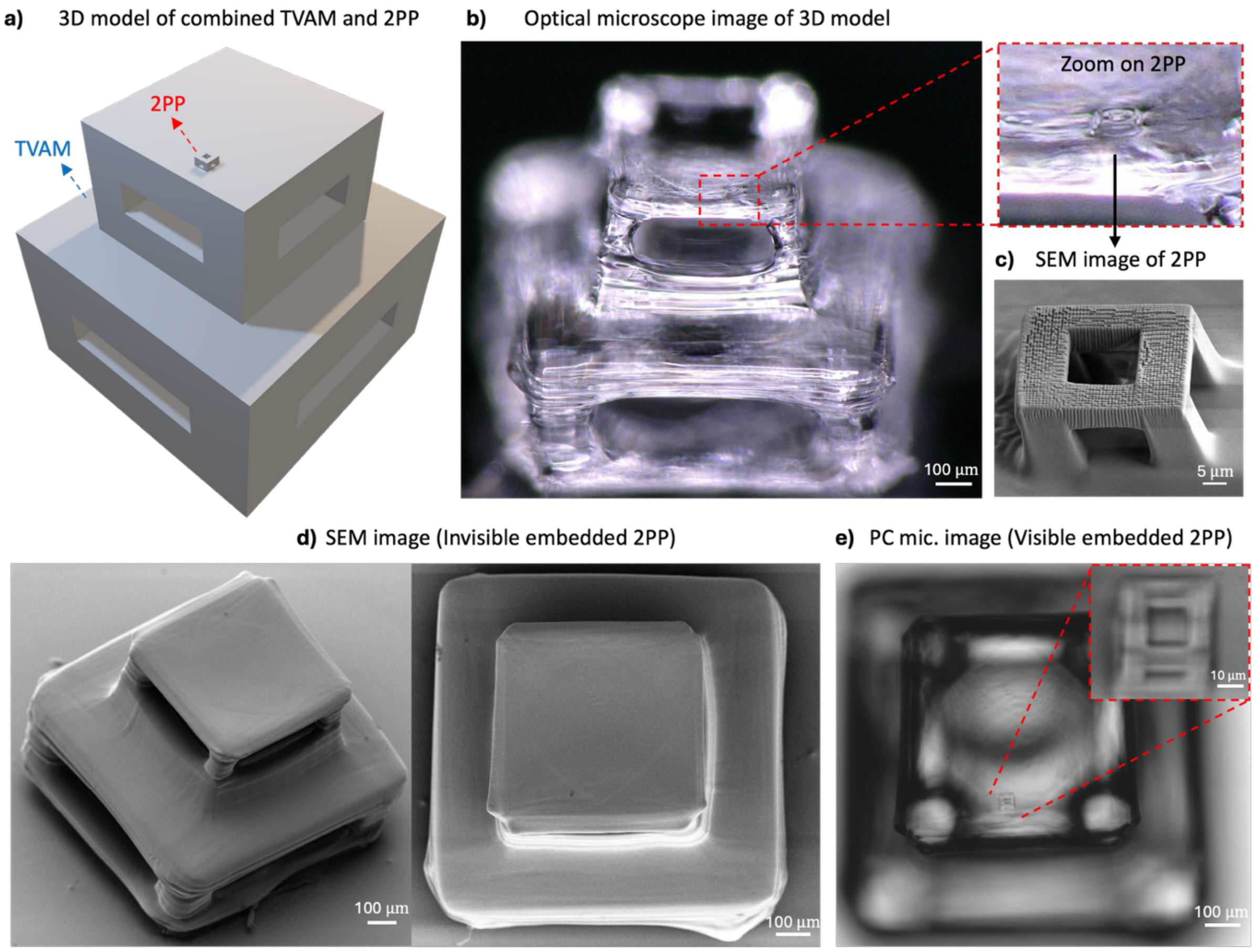

**Figure 4.** Two stacked rectangular prisms overprinted by TVAM, combined with a micrometer-scale bridge-like 2PP-printed structure. a) 3D CAD model for the combination of TVAM (two stacked rectangular prisms) and 2PP (bridge-like structure) printing. b) Digital microscope image of the printed model. The inset shows a magnified view of the bridge-like structure printed by 2PP. Scale bar is 100 µm. c) SEM image of the 2PP-printed bridge-like structure. Scale bar is 5 µm. d) SEM images of the printed final structure from a 40° tilt (left) and top view (right), showing TVAM-overprinted rectangular prisms, while the embedded 2PP-printed bridge-like structure is not visible. The scale bars are 100 µm. e) PC microscope image of the same structure in which the embedded 2PP-printed part is observable. The inset shows a magnified view of the embedded 2PP-printed bridge-like structure. The scale bars are 100 µm and 10 µm, respectively.

In this work, TVAM is employed to fabricate the overall millimeter-scale body of the structure, such as the gear and the stacked prisms, since these volumes mainly require feature sizes on the order of tens of micrometers and benefit from the rapid, high volumetric throughput of TVAM. Printing the entire millimeter-scale volume by 2PP alone would require stitching many xy-axis scanning fields, resulting in a prohibitively long fabrication time. Therefore, 2PP is used only in selected regions where sub-micrometer features or intricate geometries are needed, such as the QR-code pattern on the gear or the bridge-like structure embedded in the TVAM-printed prisms, while TVAM cannot achieve these sub-micrometer features. This division of tasks leverages the strengths of both techniques: TVAM provides fast bulk fabrication, while 2PP adds localized high-resolution functionality.

To assess our system's throughput, we experimentally measured the minimum feature size achievable with both methods. On the TVAM side, the fabrication of a 1 $mm^3$ 3D structure takes 12.6 seconds with a minimum feature size of approximately 38 µm corresponding to a printing rate of approximately $1.5\times10^3$ voxels/s. On the 2PP side, the voxel rate in our proof-of-principle results is limited to 20 voxels/s (50 ms exposure time) by the frame rate of the LCOS SLM (60 fps) and by the MATLAB control program. This minimum exposure time of 50 ms renders precise dose control difficult for deep sub-diffraction-limit feature size[32]. This is because 2PP process exhibits an intensity threshold below which no solid polymer forms[23], which is due to the nonlinear optical nature and the competition among initiation, inhibition, and molecular diffusion. At an intensity above this intensity threshold, the minimum exposure time of 50 ms leads to a dose exceeding the polymerization threshold by many times. Consequently, the minimum feature size in our 2PP printing was 830 nm.

Since our technique integrates two drastically different methods into one system, it is difficult to define an overall figure of merit that measures its combined throughput fairly. While 2PP fabrication works on the basis of voxels, there is no inherent physically defined voxel in TVAM although a conceptual equivalent can be defined. The voxel rate on either the TVAM or the 2PP system does not provide an accurate representation of the overall combined system. Here, we propose the concept of "volume feature ratio" as a general measurement of the complexity of a print, which is defined as the ratio of the volume printed by TVAM and by the volume printed via 2PP. We further define "time rate of feature ratio" as the throughput for a given technique, which is defined as the complexity measured as the feature ratio printed per unit time . The volume fabricated by 2PP is typically only a small fraction of the total TVAM volume, as its role is to introduce precise, well-defined features. A representative case is a volume fraction of 0.1 %, corresponding to a region whose linear dimensions are 10 % of those of the TVAM volume along each axis. The throughput of the combined technique is thus $(V_{TVAM}/\delta V_{2PP})/t$, where $V_{TVAM}$ is the TVAM volume, $\delta V_{2PP}$ is the 2PP printing voxel volume, and $t = t_{TVAM} + t_{2PP}$ is the total printing time, which includes time for TVAM ($t_{TVAM}$) and 2PP ($t_{2PP}$) printing. To remain consistent and comparable with the conventional definition of throughput, we keep the unit of voxels/s here. Under this measurement, our system has a volume feature ratio of $1\times10^9$ and a throughput of $2\times10^4$ voxels/s, given that $V_{TVAM} = 1\ \mathrm{mm}^3$, $\delta V_{2PP} = 1\ \mathrm{\mu m}^3$, and $t = 5 \times 10^4$ s, which is dominated by the 2PP printing to fulfill the $1\times10^{-3}$ $mm^3$ volume.

We note that both TVAM and 2PP fabrication speed can be significantly improved in a future implementation. The TVAM printing time can be reduced by up to one order of magnitude, as demonstrated in Ref.[58] for a highly optimized system. For 2PP, while a resonant galvanometer scanning system and a high-speed optical modulator represent a straightforward approach that would greatly increase the 2PP fabrication speed[59], there is a limit to the maximum intensity required for fast scanning, beyond which material damage occurs. Optimization of the photoinitiator plays a critical role in achieving the ultimate throughput. The 2PA cross-section of TPO-L currently used is only around 0.1 GM at 800 nm wavelength[44]. On the other hand, the anthracene-based photoinitiator 9,10-bis-pentyloxy-2,7-bis[2-(4-dimethylamino-phenyl)-vinyl]anthracene (BPDPA) is highly sensitive and efficient at 800 nm with a 2PA cross-section of 294 GM[60]. By replacing TPO-L with BPDPA, the required intensity to achieve the same dose could be reduced by 2940 times. With this in mind, we consider an example where a resonating galvanometer scanner is used in place of the slow SLM to achieve a scanning rate of 8 kHz in a 2.3 mm field of view in our microscope objective, leading to a bidirectional scanning speed of 18.4 m/s. Given the objective NA of 0.45, the focus spot size for 2PP is approximately 2.1 µm at 780 nm wavelength, and the 2PP-active spot size is 1.48 µm, requiring a minimum modulation bandwidth of 12 MHz for the optical modulator , which is readily achievable with commercial acousto-optic modulators. The voxel rate of the 2PP system would reach $1.2\times10^{7}$ voxels/s at an induction time of $8\times10^{-8}$ s without regard to the laser power and the intensity limit. To estimate the required average laser power $P$, we note that the same 2PP dose level of 13 mW by 50 ms can be maintained with more sensitive PI and shorter induction time while taking into account the quadratic response of 2PA to the laser intensity[32]. The equation $t_0/t = \left(\sigma_2^{(B)}/\sigma_2^{(T)}\right)(P/P_0)^2$ , where $t = 8\times10^{-8}$ s is the required induction time, $\sigma_2^{(B)} = 294\ \mathrm{GM}$ and $\sigma_2^{(T)} = 0.1\ \mathrm{GM}$ are the 2PA cross-section of BPDPA to TPO-L, respectively, and $P_0 = 13\ \mathrm{mW}$ and $t_0 = 50\ \mathrm{ms}$ is the average laser power and the induction time used in our experiment, respectively. With these values, the optical power required is $P = 0.19$ W. Thus, a femtosecond laser with a moderate power can achieve this high-throughput printing, resulting in a peak intensity of $9.6\times10^{11}$ $\mathrm{W/cm^2}$ at the focal spot (based on 70 fs pulse width and 80 MHz repetition rate), which is well below common material breakdown threshold. Moreover, the fabrication speed could be further improved with an even faster version of resonant galvanometer (for example, 24 kHz) and more sensitive photoresin formulations[61,62]. With these optimizations, an overall (TVAM + 2PP) throughput of $2.8\times10^{8}$ voxels/s could be achieved. Therefore, the advantage of the heterogeneous hybrid approach compared with a unimodal method[63] is clear, printing larger volumes at a higher throughput using lower fs laser power.

From an economic perspective, the cost of the present hybrid platform is largely driven by the fs laser source used in 2PP. Nevertheless, recent work on diode-laser-based 2PP and on two-step absorption 3D nanoprinting demonstrates that compact and lower-cost laser architectures can still achieve high-resolution nonlinear printing using monolithic diode lasers or inexpensive continuous-wave diodes with tailored two-step photoinitiators[64–66]. While not used here, such schemes are conceptually compatible with our hybrid TVAM–microfabrication framework and represent promising options for future, more economical implementations.

## 4. Conclusion

We introduced a unified multiscale 3D printer proof of concept that integrates 2PP with layer-less single-photon TVAM, bridging the gap between high-precision and the exceptional speed and the volumetric throughput of the TVAM approach. By combining the nanometer-scale precision of 2PP with the exceptional volumetric throughput of TVAM, the system bridges the resolution-throughput trade-off in light-based additive manufacturing and enables rapid fabrication of mesoscale structures spanning three orders of magnitude in length scale.

The volumetric nature of TVAM enables high throughput and rapid 3D fabrication by eliminating mechanical scanning and layer-by-layer stacking, both of which inherently limit the printing speed in conventional optical lithography. Within the unified platform, TVAM is used to rapidly fabricate millimeter-scale structures, while 2PP features the millimeter-scale TVAM-printed objects with localized nanometer-scale resolution. To support this hybrid approach, we employed Dr.TVAM framework, which accounts for ray optical effects including refraction, absorption and reflection to optimize the tomographic projection patterns. This enables accurate overprinting of TVAM-printed structures on a rotating glass rod that simultaneously serves as the build platform, enabling direct accessibility for subsequent 2PP both on the surface and within the volumetric TVAM-printed structures.
We experimentally validated the approach through several demonstrations. First, millimeter-scale gears were fabricated via TVAM in 12 seconds on the glass rod, on which, without changing the resin or applying post-processing, a QR code with 830 nm feature size was written by 2PP. The 3D printer further enabled the fabrication of multiple 2PP-printed structures along the rotational path of the rod, by either SLM scanning or the rod rotation, thereby extending the accessible FOV and enhancing overall flexibility of the multiscale 3D printer. More complex hybrid structures were also realized. Millimeter-scale rectangular prisms with lateral through-holes were first fabricated via TVAM and subsequently complemented with bridge-like 2PP-printed structures, both on the surface and embedded within the volumetrically polymerized structures, demonstrating highly localized and precise fabrication with sharp and well-defined edges. The structures demonstrated here primarily serve as proof-of-concept examples rather than fully optimized devices.
We proposed a concept of “volume feature ratio”, as a general measurement of the complexity of a print, which is defined as the ratio of the volume printed by TVAM and by the volume printed via 2PP, and its time rate as a reference standard for throughput benchmarks of the hybrid system. Under this concept, we define 1 $mm^3$ total print volume as a standard volume, in which 0.1 % of the volume (the standard 2PP ratio) is to be printed with 2PP. Under this definition, the overall throughput of our system is $2\times10^4$ voxels/s.
The performance of the system could be further improved with a number of optimizations. By using a more sensitive PI, such as BPDPA, instead of TPO-L, a fast resonant galvanometer scanner at 8 kHz, and an optimized TVAM configuration, a throughput of $2.8\times10^8$ voxels/s could in principle be achieved in a TVAM-2PP hybrid system.
This unified multiscale 3D printer opens new possibilities for applications requiring rapid, high-throughput fabrication of millimeter-scale structures with precisely localized sub-micrometer features, including bioengineered scaffolds with pore architectures at different length scales combined in a single construct, microfluidic devices that integrate microchannels and microvalves into a mechanically robust body, and integrated micro-optical components where internal waveguides, resonators and microlenses are embedded in a larger transparent substrate.

## Acknowledgments

This project has received funding from the Swiss National Science Foundation 2000-1-240074 under grant number 10007068 - “Neural precision holographic volumetric additive manufacturing”. The authors thank to Julie Anna Gabrielle Riou for contributing the 3D CAD designs.

## Author contributions

**Buse Unlu:** Writing - Original Draft, Performed experiments, Methodology, Software, Validation, Formal analysis, Investigation, Data Curation, Visualization, Project administration. **Felix Wechsler:** Writing - Review & Editing, Methodology, Software, Validation, Formal analysis, Investigation, Data Curation, Visualization. **Ye Pu:** Writing -

Review & Editing, Methodology. **Christophe Moser:** Writing - Review & Editing, Conceptualization, Methodology, Supervision, Project administration, Funding acquisition.

**Data availability**

The full configuration files and meshes for TVAM can be found at: https://github.com/EPFL-LAPD/A-unified-multiscale-3D-printer-combining-single-photon-TVAM-and-Two-Photon-Polymerization. Further data will be made available on request.

**Conflict of interests**

Christophe Moser is a shareholder in Readily3D. Buse Unlu, Ye Pu and Christophe Moser have filed a Patent (EP26152608) entitled "additive manufacturing apparatus and method". All other authors declare no conflict of interest.

# Supplementary Materials for

# A unified multiscale 3D printer combining single-photon Tomographic Volumetric Additive Manufacturing and Two-Photon Polymerization

Buse Unlu[1,*], Felix Wechsler[1], Ye Pu[1], Christophe Moser[1]

[1]Laboratory of Applied Photonics Devices, School of Engineering, Ecole Polytechnique Fédérale de Lausanne, CH-1015,
Lausanne, Switzerland

Corresponding author: buse.unlu@epfl.ch

## Experimental setup for tomographic volumetric additive manufacturing

The optical setup used for tomographic volumetric additive manufacturing (TVAM) is illustrated in Figure S1. A continuous-wave (CW) laser diode at 405 nm (HL40033G, Ushio) is coupled into a square-core multimode fiber (70 µm x 70 µm core size, NA = 0.22, WF 70×70/115/200/400N, CeramOptec) by means of a collimator (L1, f = 4.02 mm, NA = 0.6, F671APC-405, Thorlabs) and an aspheric lens (L2, f = 15.3 mm, NA = 0.16, C260TMD-A, Thorlabs), respectively. The light output from the fiber is collimated using an achromatic doublet (L3, f = 100 mm, AC254-100-A-ML, Thorlabs) and projected onto the aperture of a Digital Micromirror Device (DMD, V-7000 VIS, Vialux). The light patterns generated by the DMD are directed through mirrors (M1 and M2) and demagnified by 2 via a lens pair of achromatic doublets, L4 (f = 300 mm, AC508-300-A-ML, Thorlabs) and L5 (f = 150 mm, AC254-150-A-ML, Thorlabs). An iris placed in the Fourier plane between L4 and L5 acts as a spatial filter blocking high diffraction orders from the DMD and constraining the angular divergence of the beam.
A quartz glass vial (refractive index 1.4696, inner diameter 4.30 mm) is mounted on a motorized rotary stage (X-RSW60C, Zaber), which is then placed on top of a one-axis motorized translation stage (PT1/M-Z9, Thorlabs) providing z-axis motion. This assembly is mounted on a two-axis linear translation stage (XYT1/M, Thorlabs) for xy-axis motion and is positioned at the center of the DMD image plane.
The patterns and the printing process are monitored by a camera (acA2040-55um, Basler) using two achromatic doublets, L6 and L7 (AC254-075-A-ML, Thorlabs), and illuminated by a laser diode at 650 nm (VLM-650-01 LPT, Quarton Inc.).

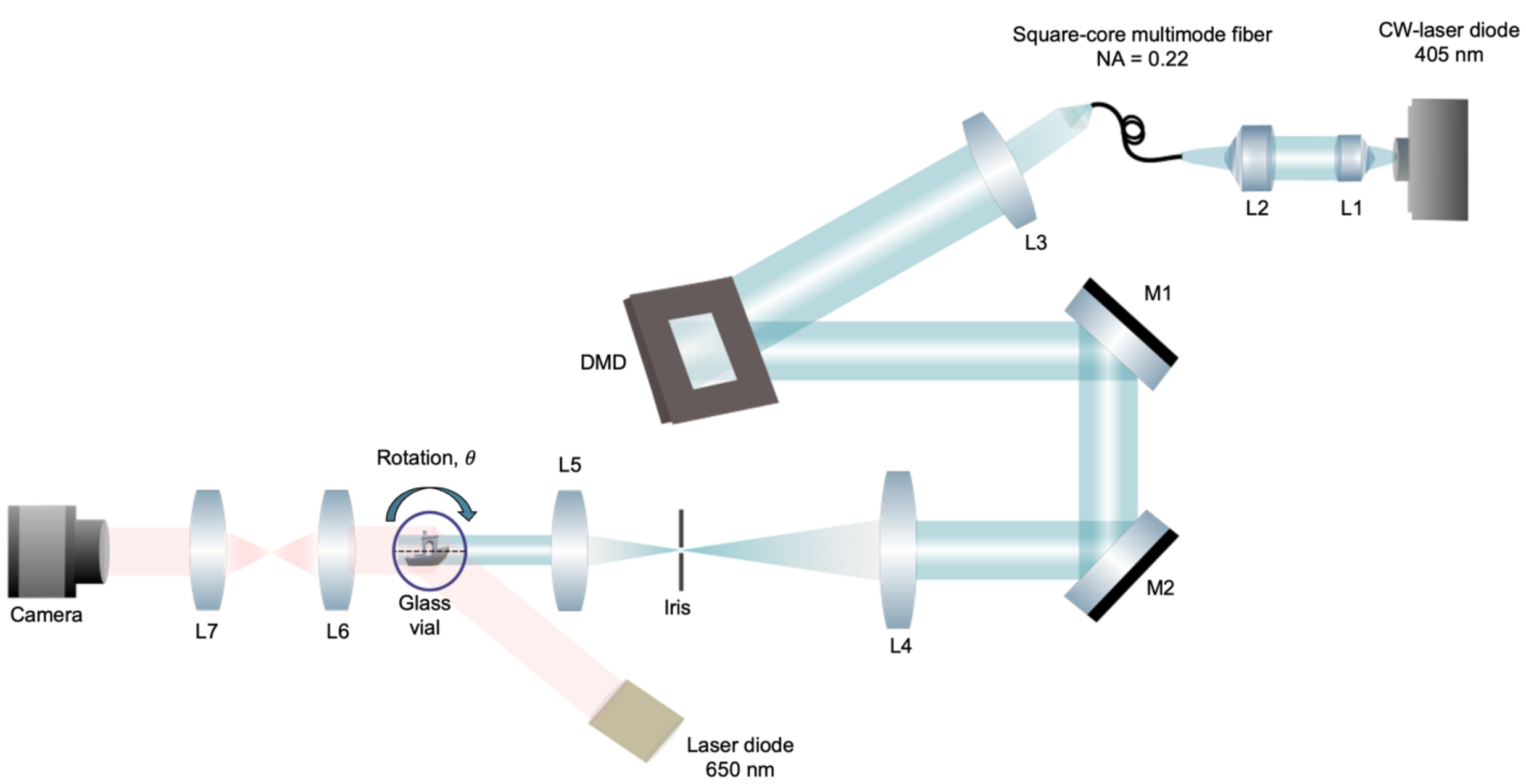

**Figure S1.** Illustration of the experimental setup for TVAM.

### Experimental setup for two-photon polymerization

The two-photon polymerization setup used in this work follows the configuration described in detail in the Supplementary Materials (Section S1 and S5) of our previous publication[1], with modifications to the printing region in Figure S2. We currently use a double-sided-cut quartz glass vial (refractive index 1.4696, inner diameter 4.30 mm) to place the photoresin. In order to satisfy the requirements of the objective lens OBL (Zeiss A-Plan 20×, NA = 0.45), a 170 µm-thick coverslip is glued to the bottom of the glass vial using an optical adhesive. A glass rod (made of borosilicate glass 3.3, ends circular cut, L = 100 ± 0.5, Diameter = 1 ± 0.05 mm, Hilgenberg GmbH) is inserted into the resin and serves as the printing platform. Both the glass vial and the glass rod are mounted on separate motorized rotary stages (X-RSW60C, Zaber) as required for TVAM (not for 2PP). Each rotary stage is placed on top of a one-axis motorized translation stage (PT1/M-Z9, Thorlabs) providing z-axis motion, and these assemblies are then mounted on a two-axis linear translation stage (XYT1/M, Thorlabs) for xy-axis motion.

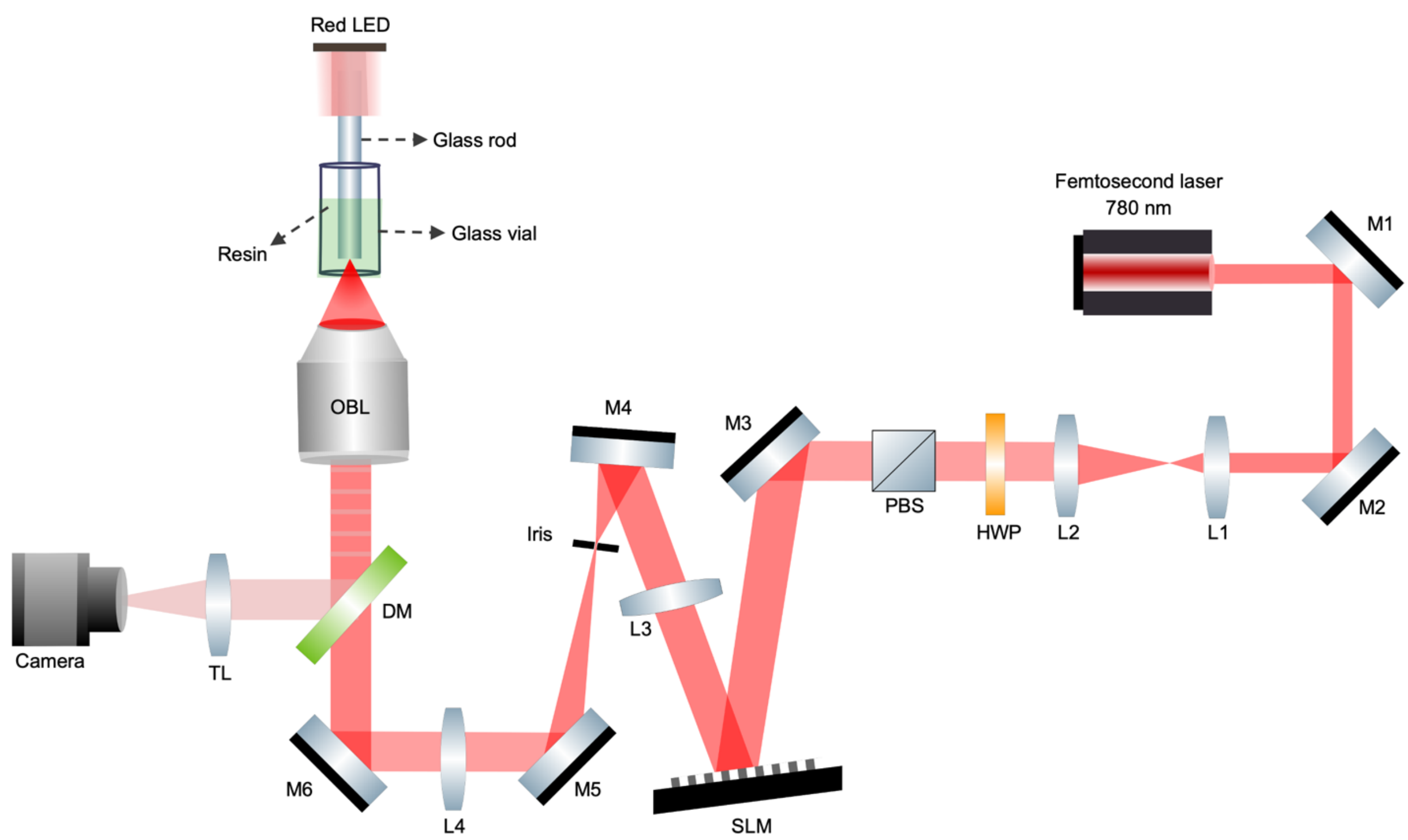

**Figure S2.** Illustration of the experimental setup for 2PP.

**Printing region of multiscale 3D Printer combining TVAM and 2PP**

A real photograph of the printing region taken from the setup where the TVAM and 2PP systems are combined is shown in Figure S3. The TVAM system (L5, in Figure S1) is positioned orthogonally to the IR fs laser beam (OBJ, in Figure S2). A glass rod is dipped into the resin vial where both are mounted on rotary stages.

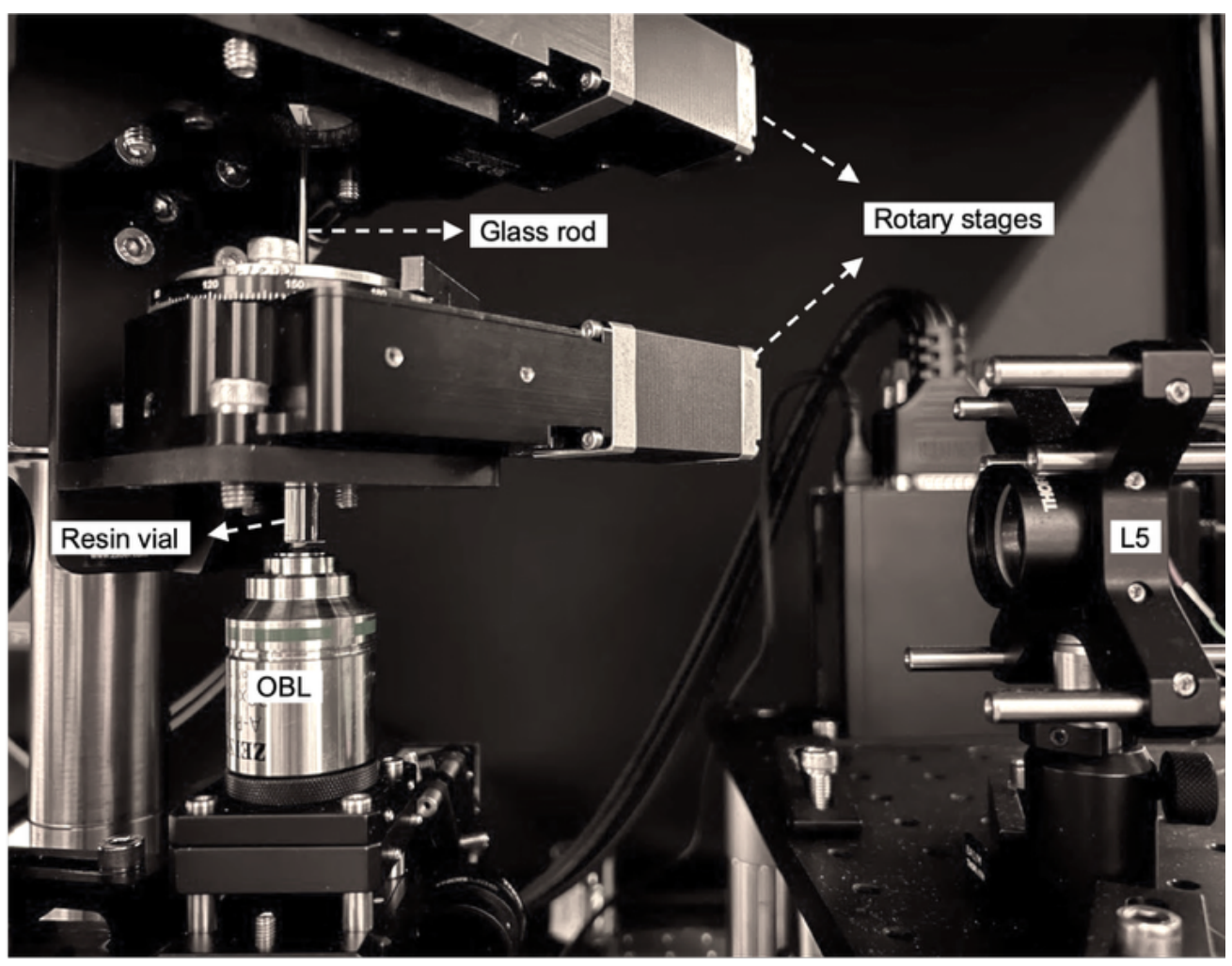

**Figure S3.** Real photograph of the printing region in the combined TVAM and 2PP setup.

**Spot diagram analysis of the tomographic volumetric additive manufacturing system**

Ray-tracing simulations were performed to check the aberration of the beam spot in the printing region of the TVAM using Zemax OpticStudio. We obtained the spot diagrams that depict the ray-intersection distribution on the image plane for three field points: on-axis (center focus) and at 10 mm and 14 mm off-axis, as shown in Figure S4, indicated by OBJ: 0.00, 10.00, and 14.00 mm. The results indicate a root-mean-square (RMS) point-spread-function (PSF) diameter of approximately 6.8 µm. Aberrations introduced by the glass vial and the photoresin were not included in these optical simulations, which are corrected in TVAM optimizations.

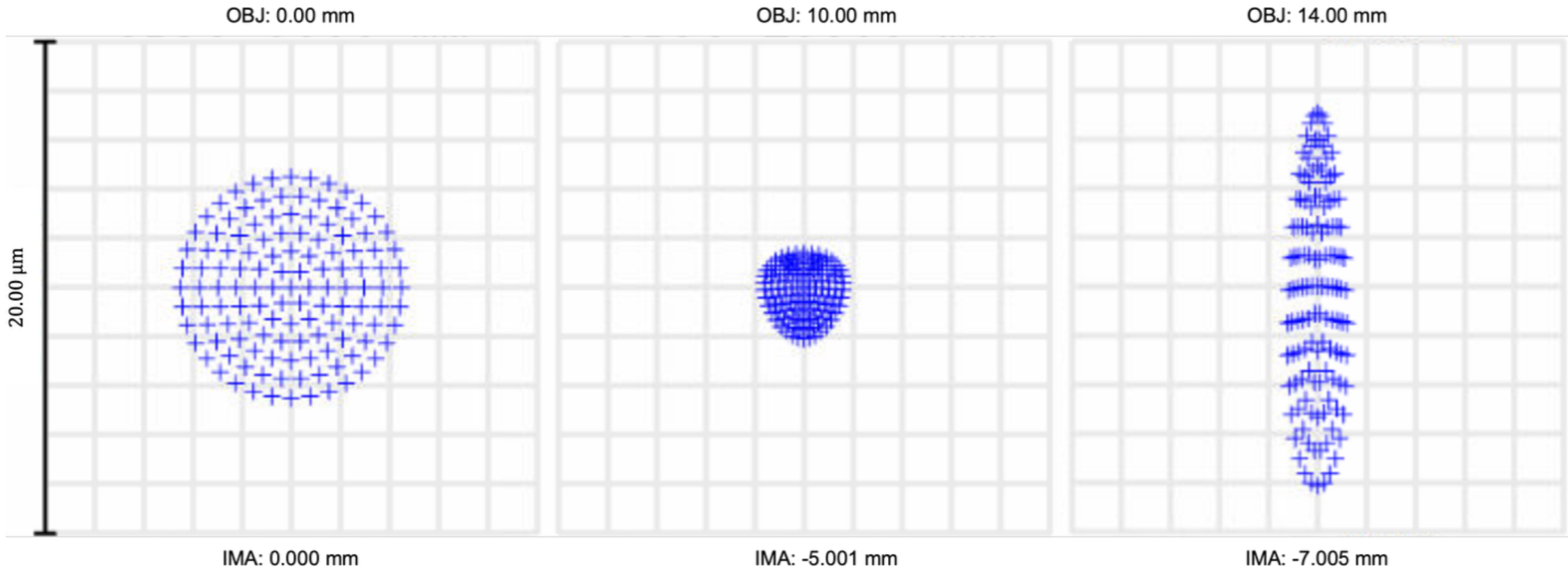

**Figure S4.** Spot diagrams from ray tracing simulations performed by Zemax OpticStudio: on-axis and off-axis at 10 mm and 14 mm, respectively.

**Photosensitive resin used for fabrication**

A photocurable resin is formulated by mixing diurethane dimethacrylate (DUDMA) and poly(ethylene glycol) diacrylate (PEGDA, average Mn 700) in a 4:1 weight ratio, with ethyl-2,4,6trimethylbenzoylphenylphosphinate (TPO-L) at a concentration of 43 mM as the photoinitiator, which absorbs both single- and two-photon excitation[1–5]. The refractive index of the resin is measured as 1.4827 at 589 nm via a refractometer (AR200, Reichert).

To characterize the optical absorption, we recorded UV–Vis spectra of the pure (as-purchased) TPO-L photoinitiator, the pure DUDMA/PEGDA resin composition without photoinitiator, and the photocurable resin containing different TPO-L concentrations (3.47, 4.42, 6.3, 7.88, 9.45, and 43 mM) shown in Figure S5a using Agilent cary 60 Uv-Vis spectrophotometer. The spectra confirm that the neat resin is essentially transparent in the visible range and that the absorption band around 400–420 nm is dominated by TPO-L. To determine the absorbance of the resin at 405 nm, we prepared samples with lower TPO-L concentrations (3.47, 4.42, 6.3, 7.88, and 9.45 mM), as the 43 mM concentration exceeds the device's measurement capabilities, and measured the absorbance at 405 nm in Figure S5b. Using these data, we applied a linear fit to extrapolate the absorbance of the resin formulated with 43 mM TPO-L, which was calculated to be 4.695, providing the extinction coefficient of 1.081 $mm^{-1}$.

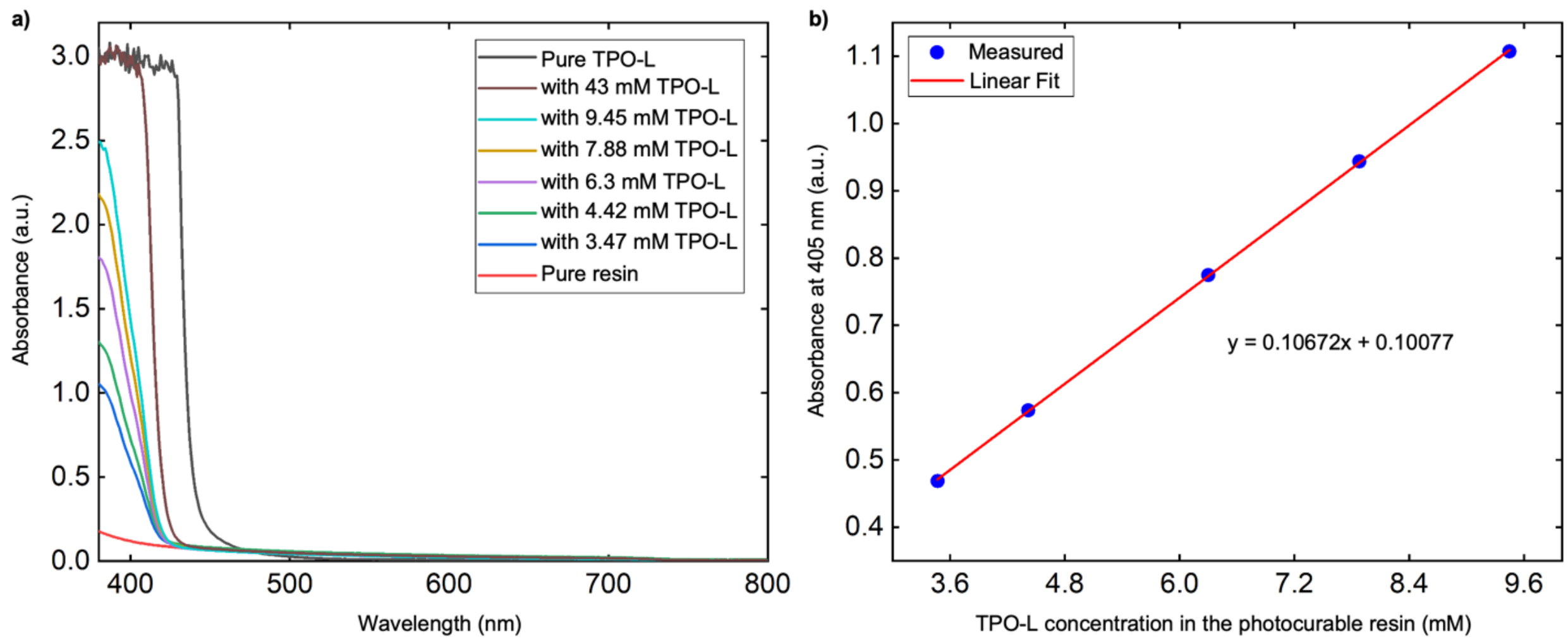

**Figure S5.** UV–Vis characterization of the photocurable resin and photoinitiator. a) Absorbance spectra of pure TPO-L photoinitiator, the pure DUDMA/PEGDA resin composition without photoinitiator, and the photocurable resin containing different TPO-L concentrations (3.47, 4.42, 6.3, 7.88, 9.45, and 43 mM). b) Absorbance at 405 nm as a function of TPO-L concentration in the resin together with a linear fit, used to extrapolate the absorbance and corresponding extinction coefficient of the 43 mM formulation.

## Energy dose calculation for TVAM

Calculating the absorbed dose of object voxels in TVAM is not a simple equation. Since in TVAM each pattern pixel projects a different intensity and since we intentionally craft the patterns to have a spatially varying absorbed 3D dose, the physical dose is heavily dependent on object properties but also resin parameters and optimization parameters (such as the thresholds and the weight of the sparsity term).

With our framework Dr.TVAM, as a first step we can calculate the energy efficiency of the patterns with

$$\text{Pattern Energy Efficiency} = \frac{\frac{\sum_{\mathrm{p} \in \mathrm{patterns}} \mathrm{p}}{\max(\mathrm{patterns})}}{\mathrm{N_{pixels}}}$$

We need to divide each pattern value by the maximum value of all pattern pixels, since the DMD in experiments normalizes all patterns with respect to the brightest value. The energy efficiency describes the ratio of projected energy and theoretically available energy.

From Dr.TVAM we can then numerically calculate how much of this projected energy is absorbed in the object voxels, how much is absorbed in the void voxels, and how much is transmitted through the vial.

Dr.TVAM exports a unitless energy per volume value with the final patterns. This value has to be multiplied by the measured power of the laser (if all DMD pixels are fully turned on) and the total printing time. The result is an energy dose in Joule per cubic meter.

For our experiments (ring-shaped micropillar array, gear and rectangular prisms), the values are, respectively,

$$73272.656\ \mathrm{m^{-3}} \times\ 73\ \mathrm{mW}\ \times\ 12.6\ \mathrm{s} =\ 67.4\ \mathrm{mJ\ cm^{-3}}$$

$$103650.523\ \mathrm{m^{-3}} \times\ 36.4\ \mathrm{mW}\ \times\ 12\ \mathrm{s} =\ 45.3\ \mathrm{mJ\ cm^{-3}}$$

and

$$60606.586\ \mathrm{m^{-3}} \times\ 70.5\ \mathrm{mW}\ \times\ 12\ \mathrm{s} =\ 51.3\ \mathrm{mJ\ cm^{-3}}$$

This value indicates the energy required to cause polymerization. Note that this energy threshold is also dependent on parameters such as concentration and type of photoinitiator, the resin, and the concentration of inhibitors. Since we use the same resin for all experiments, we also expect to have roughly the same energy threshold.

**Feature size analysis of 2PP system**

Figure S6 illustrates the quantitative analysis of the smallest printable features in the 2PP-fabricated QR code. In Figure S6a, the original digital QR code, the corresponding SEM image of the 2PP-printed QR code, and a superimposed SEM image with the QR code outline are shown to visually compare the designed pattern with the printed structure and to identify resolved features. Green regions in the overlap indicate empty spaces where no voxels were formed, helping distinguish printed voxels from background and guiding the selection of features for measurement. For the quantitative study, only features that were isolated in at least one lateral direction were selected so that their size could be measured unambiguously along that isolated dimension, resulting in a dataset of 40 individual features. The result of the feature analysis is presented as a histogram of the measured size in Figure S6b, revealing an average feature size of 830 nm with a standard deviation of 140 nm.

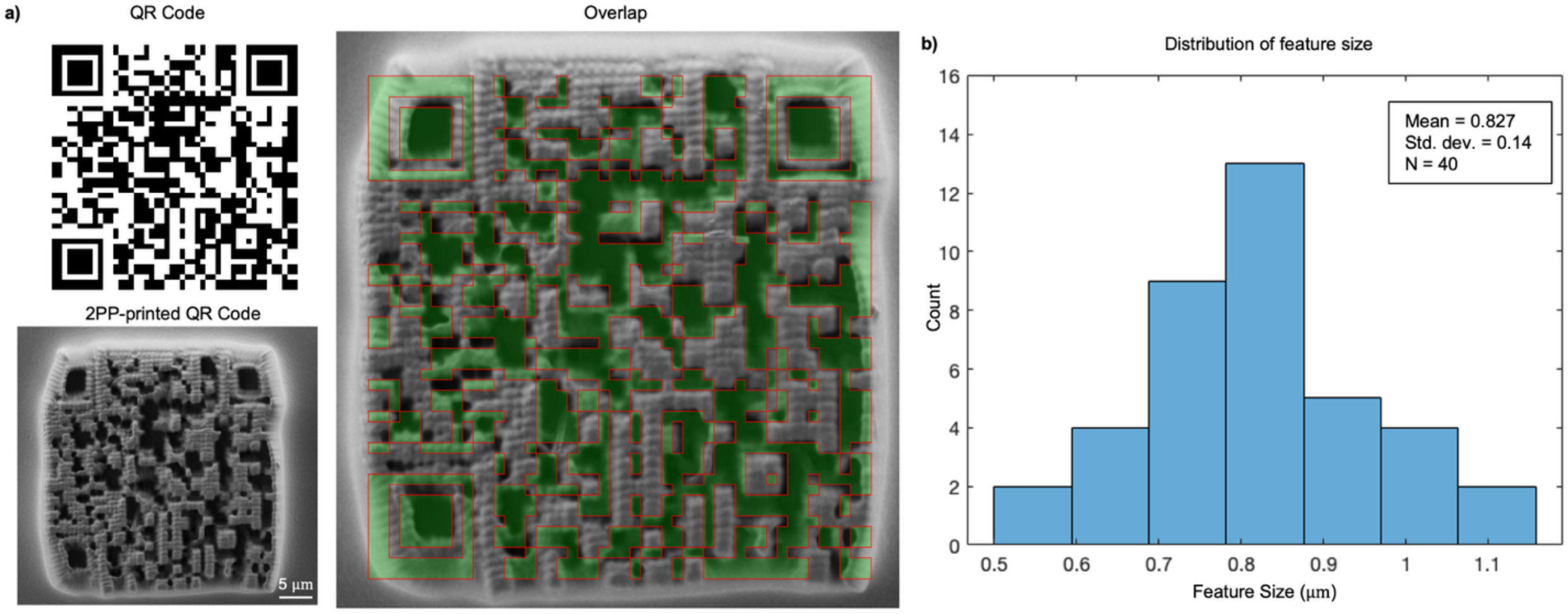

**Figure S6.** Analysis of feature size in the 2PP-printed QR code. a) The original QR code image (left top), the SEM image of the 2PP-printed QR code (left bottom), and the SEM image superposed with the QR code outline for assistance of analysis, where green area corresponds to empty space (no voxels). Only features that are isolated in at least one dimension were picked and analyzed in the isolated dimension. In total, 40 features were analyzed that satisfy this requirement. b) Result of the feature analysis as histogram of the measured size, showing an average of 0.83 μm with a standard deviation of 0.14 μm.

**Feature size analysis of TVAM system**

To address the achievable minimum size of TVAM system, we designed a ring-shaped micropillar array in which the pillar heights were 600 μm and the diameters were symmetrically varied between 500 and 20 μm within the single structure, as shown in Figure S7a. The optimized 2D patterns were projected onto the photocurable resin with a light dose of 67.4 mJ cm$^{-3}$ (the calculation is provided in the section of "Energy dose calculation for TVAM" in Supplementary Materials) and the ring-shaped micropillar array structure was printed within 12.6 seconds. Digital microscope images (VHX-5000, Keyence) acquired from different viewpoints are shown in Figure S7b–d, illustrating the overall morphology of the printed structure. As highlighted in Figure S7b, pillars with diameters down to 40 μm can be formed, whereas thinner targeted features (30, 25, and 20 μm) fail to polymerize under the same exposure conditions (Figure S7c,d).

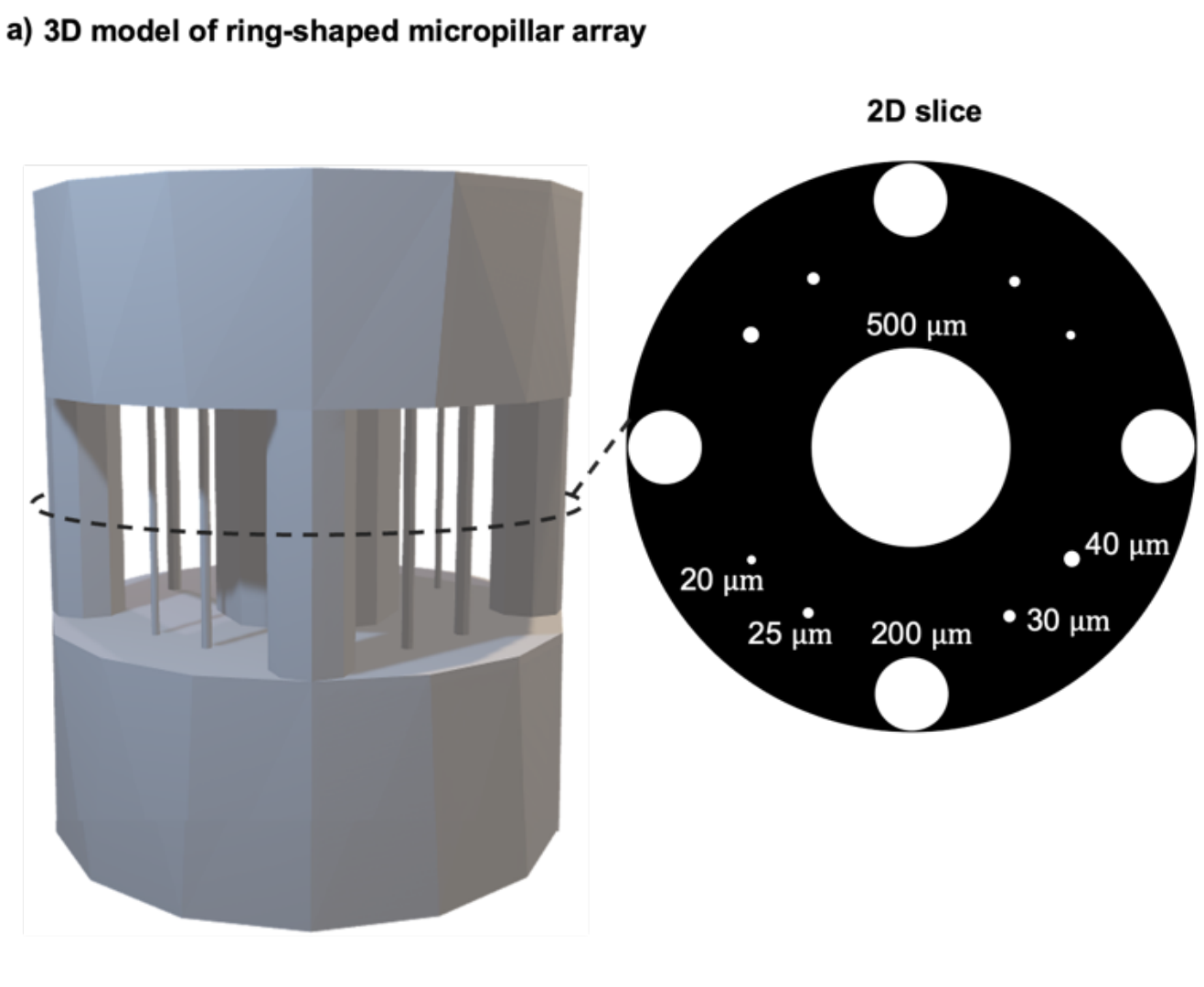

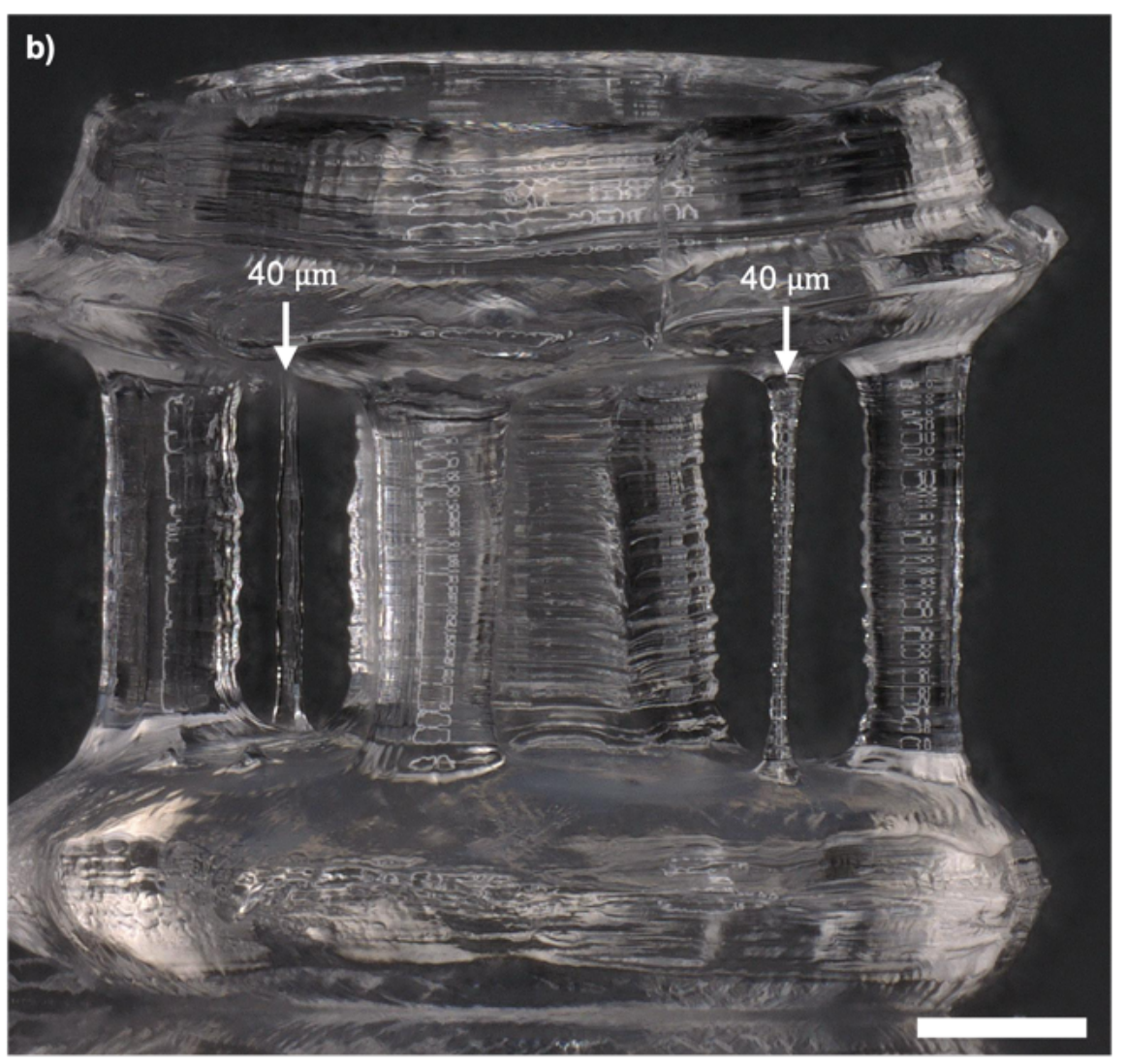

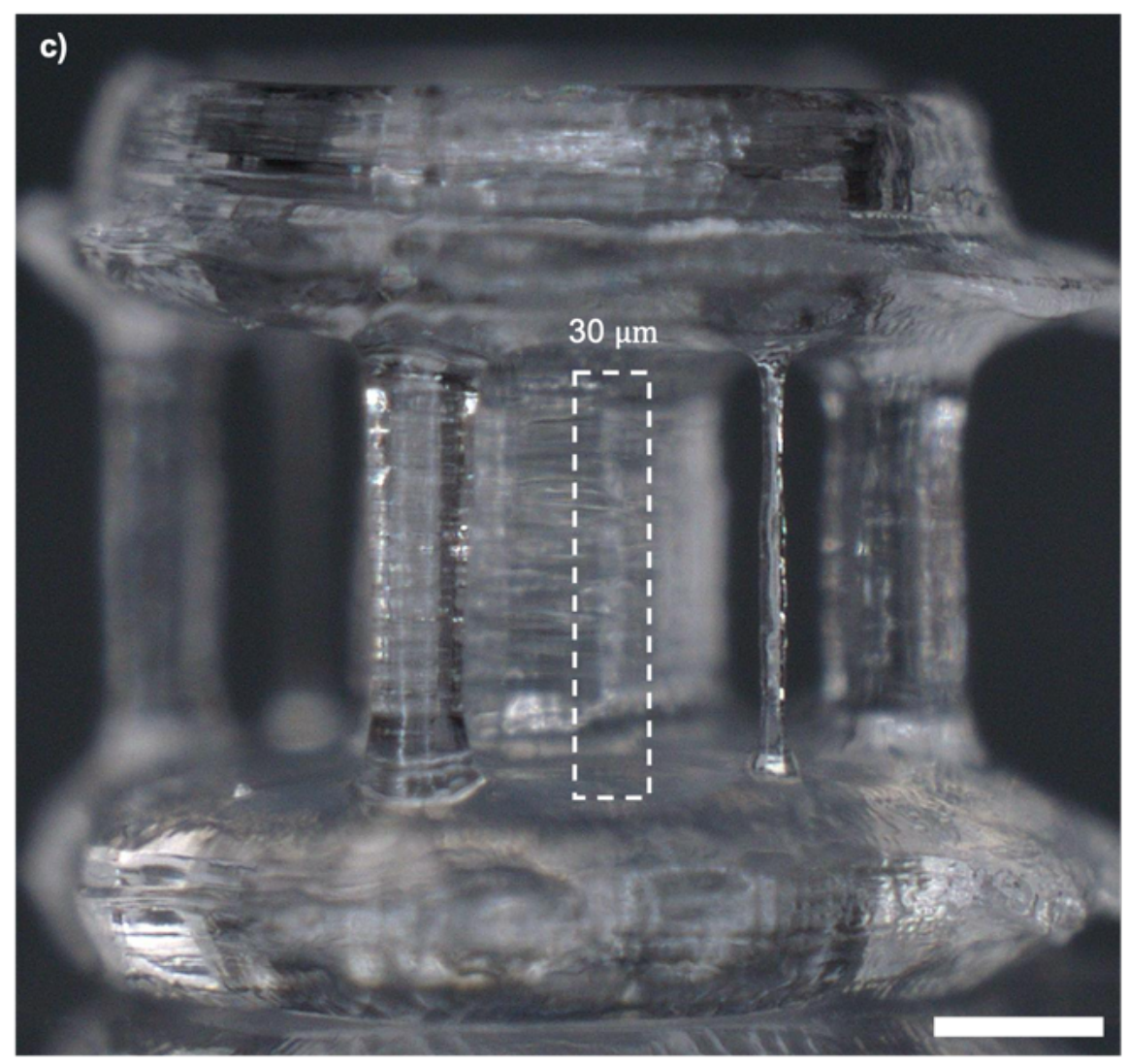

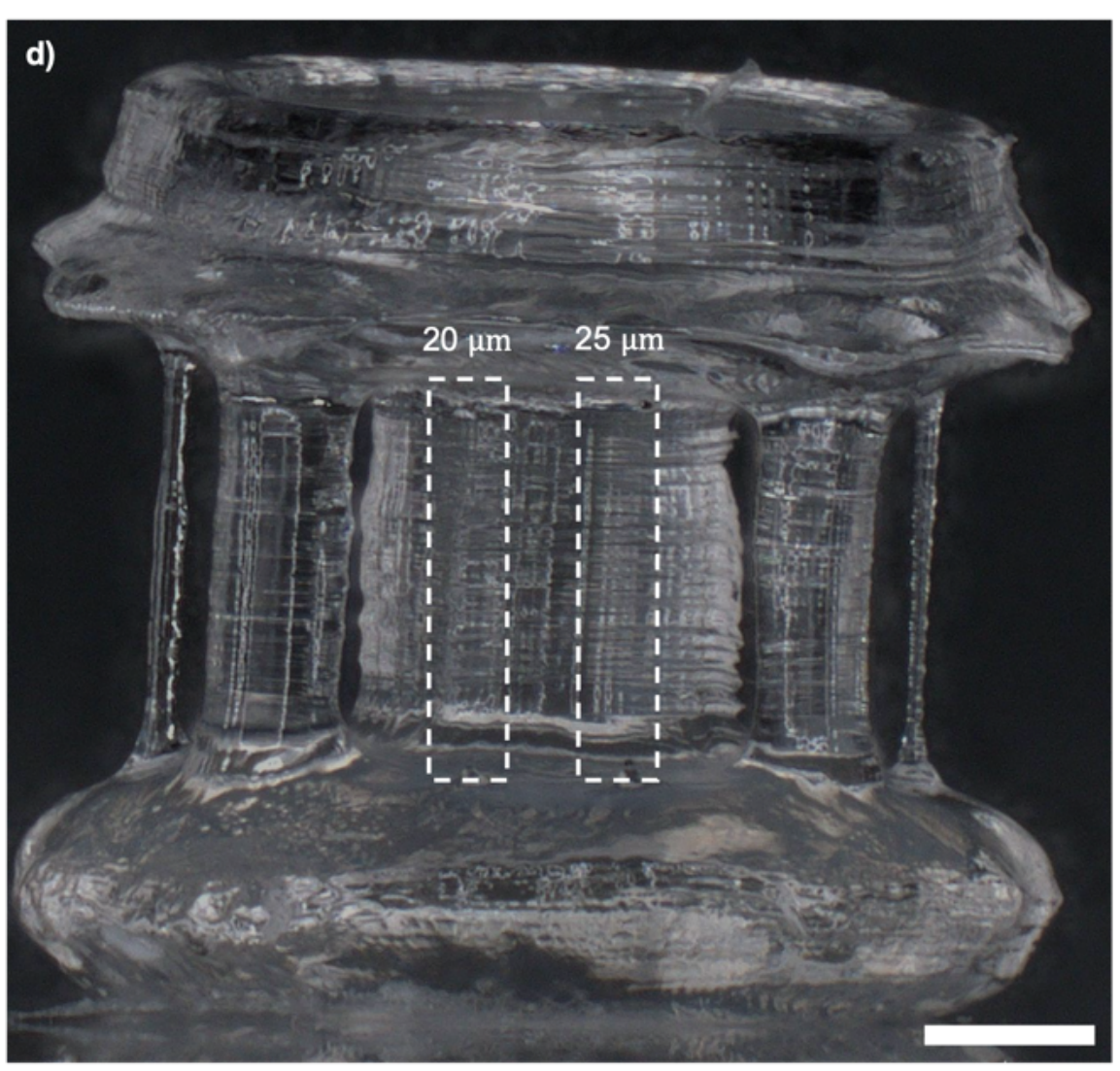

**Figure S7.** Investigation of minimum feature size achievable with the TVAM system. a) 3D model of the ring-shaped micropillar array in which the pillar diameters are symmetrically varied between 500 µm and 20 µm within a single structure shown in the 2D slice. b-d) Digital microscope images of the printed structure acquired from different viewpoints, illustrating the overall morphology of the TVAM-fabricated array. In b), white arrows indicate pillars with a designed diameter of 40 µm that are successfully formed. In c) and d), dashed rectangles highlight the regions where thinner targeted features (30, 25, and 20 µm) were designed but fail to polymerize under the same exposure conditions. Scale bar is 250 µm.

To further quantify the feature size achieved by TVAM under these fixed exposure conditions, we fabricated several identical ring-shaped micropillar arrays and measured the diameters of individual pillars (designed to be 40 µm) from different viewing directions. Representative microscope images of five pillars (A–E) are shown in Figure S8a. For each pillar, we extracted the mean diameter and standard deviation along the pillar height, which are reported in the bar chart in Figure S8b. Aggregating all measurements into a single histogram (Figure S8c) yields a distribution of pillar diameters centered at 38.3 µm with a standard deviation of 9.37 µm (N = 329). The size difference arises due to oxygen inhibition, which is more pronounced for smaller features where oxygen diffusion and quenching of radicals reduce the local degree of cure and effective pillar diameter[6–8].

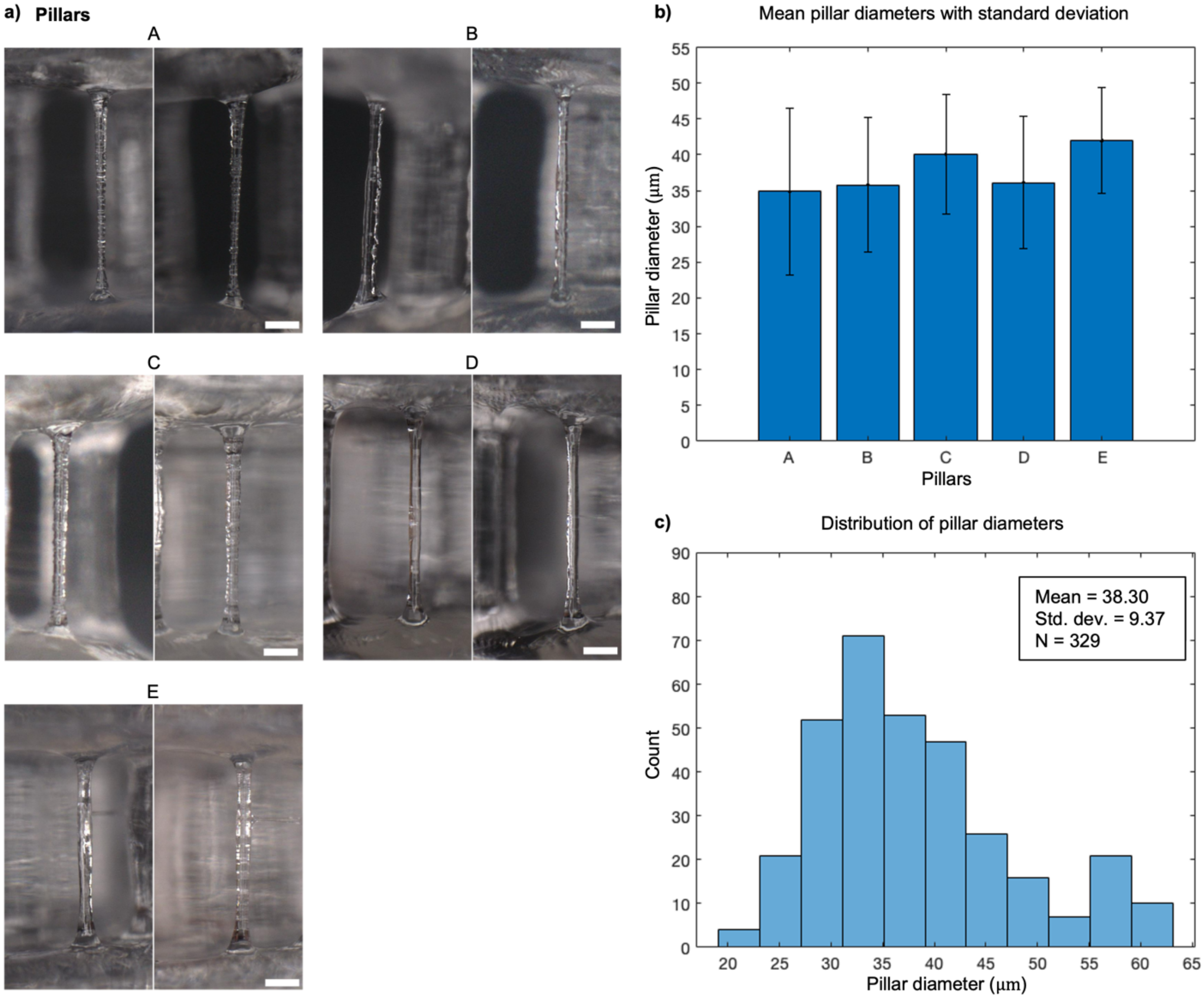

**Figure S8.** a) Digital microscope images of five pillars (A–E) from TVAM-printed ring-shaped micropillar arrays under identical exposure conditions. For each pillar, two imaging orientations are shown: a front view and a side view. b) Bar chart reporting the mean pillar diameter and standard deviation for each pillar A–E, obtained by measuring the diameter at multiple positions along the pillar height in both front and side views. c) Histogram of all measured pillar diameters (N = 329), yielding a distribution centered at 38.3 µm with a standard deviation of 9.37 µm. Scale bar is 100 µm.

## Fluid dynamics of photoresin under rotational motion

When we printed a gear with only one full rotation of the glass vial before illuminating the DMD patterns, the gear teeth were not straight but instead curved in one direction, as shown in Figure S9a. This effect arose from a mismatch between the rotational speed of the vial and the liquid resin. At the beginning of the rotation, the bulk of the liquid remains still due to the inertia. Only the liquid layer immediately next to the wall of the container moves with the container due to the nonslip condition, forming a radial velocity gradient which results in a shear stress in the liquid. The rest of the liquid is then driven by the shearing force, forming a typical shear-driven flow system[9]. Depending on the viscosity of the liquid, there is a time delay for the entire liquid bulk to rotate at the same speed before which the existence of a shear-driven flow in the liquid bulk causes printing distortion. When we rotated the vial three full rotations before illuminating the patterns, the resulting teeth were straight (in Figure S9b), as this allowed the resin to reach a uniform rotational speed throughout.

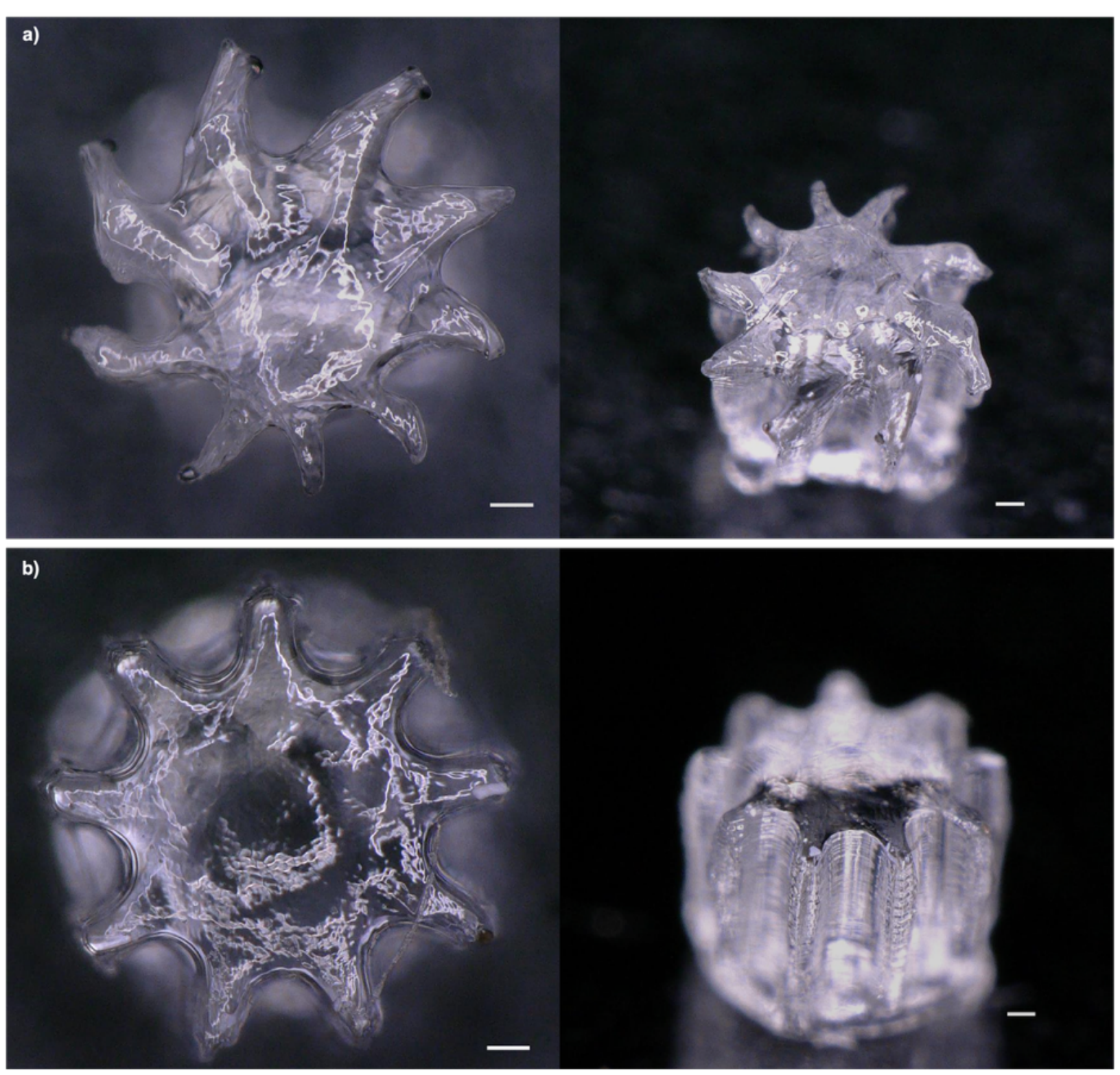

**Figure S9.** Digital microscope image of the TVAM-overprinted gear (top and side views, respectively) with a) one and b) three full rotations of the glass vial prior to DMD pattern illumination. The scale bars are 100 µm.

**Overview of hybrid approaches for 3D mesoscale printing**

Multiphoton printing offers the smallest feature sizes, reaching several tens of nanometers. The technique is well established, with commercial systems available from companies such as Nanoscribe, UpNano, and Multiphoton Optics[10]. However, because microscope objectives typically provide a limited field of view (tens to hundreds of micrometers), centimeter-scale lateral printing requires mechanical 2D stages to stitch together multiple fields of view to form a larger structure. As a result, total printing time scales with the number of stitched fields, and fabrication times on the order of hours are commonly reported. Even stitch-less mesoscale implementations require synchronized galvo and stage scanners and still take hours to fabricate centimeter-scale structures[11].
Another approach relies on micro-stereolithography (µSLA), a single-photon stereolithography technique that uses a microscope objective to achieve micrometer-scale features. In this case as well, mechanical 2D scanning or lens arrays are used to enlarge the printable surface area[12,13].

Reported fabrication times range from approximately 0.5 hours to several hours. Digital Light Processing (DLP), capable of producing tens-of-micrometer features over build areas of several square centimeters, has also been combined with µSLA to produce mesoscale structures[14], with printing times typically ranging from tens of minutes to hours.
Systems combining single-photon lithography techniques (such as DLP or SLA) with two-photon lithography remain relatively rare[15–17]. In Ref.[15], two-photon ablation, rather than additive fabrication, is used to produce mesoscale structures in 66 minutes. Refs.[16,17] correspond to patents and press releases for which, to our knowledge, no experimental results have been reported.
Other reported implementations achieve mesoscale fabrication using combinations of these techniques but employ sequential platforms: a first structure is printed on one system (e.g., DLP), after which the part is transferred to another system (e.g., two-photon printing) for further processing[18–21]. All reported approaches rely on the layer-by-layer combination of conventional lithography methods.

**Supplementary Table 1.** Summary of hybrid approaches for 3D mesoscale printing.

| Technique | Resolution | Size | Printing time | Ref. |
| --- | --- | --- | --- | --- |
| **Mesoscale printing in single platform** | | | | |
| • Two-photon<br>• Voxel-by-voxel<br>• 2D scanning by linear stage (long) + galvanometric scanning (short) | Sub-µm | cm | 2.5 hours | 11 |
| • Single-photon<br>• Layer-by-layer<br>• Microstereolithography + scanning<br>- | µm | cm | Several hours | 12 |
| • Single-photon<br>• Layer-by-layer<br>• Microstereolithography + microlens array | µm | cm | Tens of minutes | 13 |
| • Single-photon<br>• Layer-by-layer<br>• Combination of DLP (coarse) and SLA (fine) | µm | cm | Several minutes to hours | 14 |

| • Layer by layer<br>• Single-photon DLP<br>• Two-photon ablation | Sub-µm | cm | 66 minutes | 15 |
| --- | --- | --- | --- | --- |
| • Layer by layer<br>• Single-photon DLP<br>• Two-photon printing | Sub-µm | cm | No result (patent)[13],<br>No result (press release)[14] | 16,17 |
| **Mesoscale printing in separate platforms** | | | | |
| • Microstereolithography (fine)<br>• micro-milling (coarse) | µm | cm | Several hours (microfluidic chips) | 18 |
| • Layer by layer<br>• Single-photon DLP printer<br>• two photon printer | Sub-µm | cm | Minutes to hours (greenbody for ceramics) | 19 |
| • Layer by layer<br>• Single-photon LCD printer<br>• two photon printer | Sub-µm | cm | Several hours (microfluidic chips) | 20 |
| • Two-photon<br>• Layer-by-layer + batch replication<br>• Combination of 2GL, Electroplating and injection Molding | Sub-µm | cm | Several hours | 21 |